\documentclass[aps, prl, reprint, superscriptaddress]{revtex4-1} 
\usepackage{graphicx}
\usepackage{dcolumn}
\usepackage{amsmath}
\usepackage{amssymb}
\usepackage{color}
\usepackage{bm}
\usepackage{braket}
\usepackage{txfonts}

\begin{document}
\title{Liouvillian Skin Effect: Slowing Down of Relaxation Processes without Gap Closing}
\author{Taiki Haga}
\email[]{haga@pe.osakafu-u.ac.jp}
\affiliation{Department of Physics and Electronics, Osaka Prefecture University, Sakai-shi, Osaka 599-8531, Japan}
\affiliation{Department of Physics, University of Tokyo, 7-3-1 Hongo, Bunkyo-ku, Tokyo 113-0033, Japan}
\author{Masaya Nakagawa}
\affiliation{Department of Physics, University of Tokyo, 7-3-1 Hongo, Bunkyo-ku, Tokyo 113-0033, Japan}
\author{Ryusuke Hamazaki}
\affiliation{Department of Physics, University of Tokyo, 7-3-1 Hongo, Bunkyo-ku, Tokyo 113-0033, Japan}
\affiliation{Nonequilibrium Quantum Statistical Mechanics RIKEN Hakubi Research Team, RIKEN Cluster for Pioneering Research (CPR), RIKEN iTHEMS, Wako, Saitama 351-0198, Japan}
\author{Masahito Ueda}
\affiliation{Department of Physics, University of Tokyo, 7-3-1 Hongo, Bunkyo-ku, Tokyo 113-0033, Japan}
\affiliation{RIKEN Center for Emergent Matter Science (CEMS), Wako, Saitama 351-0198, Japan}
\affiliation{Institute for Physics of Intelligence, University of Tokyo, 7-3-1 Hongo, Bunkyo-ku, Tokyo 113-0033, Japan}
\date{\today}

\begin{abstract}
It is highly nontrivial to what extent we can deduce the relaxation behavior of a quantum dissipative system from the spectral gap of the Liouvillian that governs the time evolution of the density matrix.
We investigate the relaxation processes of a quantum dissipative system that exhibits the Liouvillian skin effect, which means that the eigenmodes of the Liouvillian are localized exponentially close to the boundary of the system, and find that the timescale for the system to reach a steady state depends not only on the Liouvillian gap $\Delta$ but also on the localization length $\xi$ of the eigenmodes.
In particular, we show that the longest relaxation time $\tau$ that is maximized over initial states and local observables is given by $\tau \sim \Delta^{-1}(1+L/\xi)$ with $L$ being the system size.
This implies that the longest relaxation time can diverge for $L \to \infty$ without gap closing.
\end{abstract}

\maketitle

{\em Introduction.--}
Relaxation processes of quantum systems coupled to environments are among the most fundamental nonequilibrium phenomena in condensed matter physics.
Recent experimental advances in atomic, molecular, and optical (AMO) systems provide a highly controllable platform to study dissipative dynamics of various quantum systems \cite{Bloch-08, Diehl-08, Weimer-08, Tomadin-11, Lee-11, Lee-11, Ludwig-13}.
A crucial question here is what determines the timescale for a given system to reach a steady state.
This problem is of great relevance to applications because controlling the relaxation time of a quantum system is a key to quantum control and information processing \cite{Pellizzari-95, Jaksch-00, Balasubramanian-09, Isenhower-10, Weimer-10, Lanyon-11}.

The spectral gap plays a crucial role in characterizing the relaxation time.
In isolated quantum systems, the spectral gap of a Hamiltonian determines the timescale of low-energy excitations.
The timescale diverges at a critical point where the spectral gap closes \cite{Sachdev}.
In quantum dissipative systems, the relaxation dynamics is characterized by the eigenspectrum and eigenmodes of a Liouvillian that governs the time evolution of the density matrix.
The Liouvillian spectral gap (or the asymptotic decay rate) $\Delta$ is defined as the smallest modulus of the real part of nonzero eigenvalues.
It has been postulated in many studies that the longest relaxation timescale of a quantum dissipative system is given by $\Delta^{-1}$ \cite{Cai-13, Bonnes-14, Znidaric-15}.
It is also known that in dissipative phase transitions the closing of the Liouvillian gap results in the divergence of the relaxation timescale \cite{Kessler-12, Honing-12, Horstmann-13, Minganti-18}.
However, while the Liouvillian gap characterizes asymptotic convergence to a steady state, transient behavior cannot generally be inferred from the gap alone.
In fact, significant differences are known to arise between the spectral gap and the mixing time, which are called cutoff phenomena \cite{Levin, Diaconis-96, Kastoryano-12, Kastoryano-13, Vernier-20}.
It is therefore natural to ask whether there is a general relationship that links transient relaxation behavior not only to the spectral gap but also to other properties of the Liouvillian.

In this Letter, we present a general relationship among the relaxation time, the Liouvillian gap, and the spatial structure of Liouvillian eigenmodes for quantum dissipative systems in which some eigenmodes are exponentially localized near a boundary of the system.
In dissipative systems driven out of equilibrium, localization phenomena of excitation modes near boundaries have attracted much attention in the context of the bulk-edge correspondence in non-Hermitian topological matter, known as the non-Hermitian skin effect \cite{Lee-16, Gong-18, Yao-18, Kunst-18, Thomale-18, Song-19, Borgnia-20, Okuma-20, Okugawa-20, Kawabata-20}.
Here, we refer to the localization of Liouvillian eigenmodes as the Liouvillian skin effect to emphasize that our study concerns Liouvillian spectra.
We derive a relation between the maximal relaxation time, the Liouvillian gap, and the localization length of eigenmodes (see Eq.~(\ref{tau_Delta_xi})).
We also propose a prototypical asymmetric-hopping model that exhibits the Liouvillian skin effect.

{\em Liouvillian skin effect and relaxation time.--}
Within the Born-Markov approximation \cite{Daley-14, Sieberer-16}, the time evolution of the density matrix $\hat{\rho}$ is described by a master equation \cite{Lindblad-76, Gorini-76},
\begin{equation}
\frac{d\hat{\rho}}{dt} = \mathcal{L}(\hat{\rho}) := -i[\hat{H},\hat{\rho}] + \sum_{\alpha} \left( \hat{L}_{\alpha} \hat{\rho} \hat{L}_{\alpha}^{\dag} - \frac{1}{2} \{ \hat{L}_{\alpha}^{\dag} \hat{L}_{\alpha}, \hat{\rho} \} \right),
\label{Liouvillian}
\end{equation}
where $\hat{H}$ is the Hamiltonian of the system, $\hat{L}_{\alpha}$ is the Lindblad operator, $[\hat{A},\hat{B}]\equiv\hat{A}\hat{B}-\hat{B}\hat{A}$, and $\{\hat{A},\hat{B}\}\equiv\hat{A}\hat{B}+\hat{B}\hat{A}$.
The Planck constant $\hbar$ is set to unity throughout this Letter.
The Born-Markov approximation is known to be justified for typical AMO systems such as trapped two-level atoms with spontaneous emission and an optical cavity with photon loss \cite{Daley-14, Sieberer-16}.
We denote the dimension of the Hilbert space of the system as $D$.
With the inner product $(\hat{A}|\hat{B}):=\mathrm{Tr}[\hat{A}^{\dag}\hat{B}]$, a set of operators forms a $D^2$-dimensional Hilbert space.
The right and left eigenmodes of the Liouvillian $\mathcal{L}$ are defined by
\begin{equation}
\mathcal{L}(\hat{\rho}^{\mathrm{R}}_j) = \lambda_j \hat{\rho}^{\mathrm{R}}_j, \:\: \mathcal{L}^{\dag}(\hat{\rho}^{\mathrm{L}}_j) = \lambda_j^* \hat{\rho}^{\mathrm{L}}_j, \:\: (j=0,1,...,D^2-1),
\end{equation}
where $\lambda_j$ is the $j$th eigenvalue and $\mathcal{L}^{\dag}$ is given by
\begin{equation}
\mathcal{L}^{\dag}(\hat{\rho}) = -i[\hat{\rho},\hat{H}] + \sum_{\alpha} \left( \hat{L}_{\alpha}^{\dag} \hat{\rho} \hat{L}_{\alpha} - \frac{1}{2} \{ \hat{L}_{\alpha}^{\dag} \hat{L}_{\alpha}, \hat{\rho} \} \right).
\label{Liouvillian_dag}
\end{equation}
The steady state $\hat{\rho}_{\mathrm{ss}}$ is the right eigenmode corresponding to the zero eigenvalue, $\mathcal{L}(\hat{\rho}_{\mathrm{ss}})=0$, and we set $\hat{\rho}^{\mathrm{R}}_0=\hat{\rho}_{\mathrm{ss}}$.
Suppose that all eigenvalues are arranged in descending order of their real parts: $0=\mathrm{Re}[\lambda_0]>\mathrm{Re}[\lambda_1]\geq...\geq\mathrm{Re}[\lambda_{D^2-1}]$.
Each eigenmode is normalized as $\|\hat{\rho}^{\mathrm{R}}_j\|_{\mathrm{tr}}=\|\hat{\rho}^{\mathrm{L}}_j\|_{\mathrm{tr}}=1$, where $\|\hat{A}\|_{\mathrm{tr}}:=\mathrm{Tr}[(\hat{A}^{\dag}\hat{A})^{1/2}]$.
The right and left eigenmodes corresponding to different eigenvalues are orthogonal to each other: $(\hat{\rho}^{\mathrm{L}}_j|\hat{\rho}^{\mathrm{R}}_k)=0 \ (\lambda_j\neq\lambda_k)$.

An arbitrary initial state $\hat{\rho}_{\mathrm{ini}}$ can be expanded as
\begin{equation}
\hat{\rho}_{\mathrm{ini}} = \hat{\rho}_{\mathrm{ss}} + \sum_{j=1}^{D^2-1} c_j \hat{\rho}^{\mathrm{R}}_j,
\label{rho_expansion_ini}
\end{equation}
where $c_j$ is written as
\begin{equation}
c_j= \frac{ (\hat{\rho}^{\mathrm{L}}_j|\hat{\rho}_{\mathrm{ini}}) }{ (\hat{\rho}^{\mathrm{L}}_j|\hat{\rho}^{\mathrm{R}}_j) }.
\label{rho_expansion_coefficients}
\end{equation}
The time evolution of the density matrix is given by
\begin{equation}
\hat{\rho}(t) = \hat{\rho}_{\mathrm{ss}} + \sum_{j=1}^{D^2-1} c_j e^{\lambda_j t} \hat{\rho}^{\mathrm{R}}_j.
\label{rho_expansion_t}
\end{equation}
The Liouvillian gap is defined by $\Delta=|\mathrm{Re}[\lambda_1]|$, which is also called the asymptotic decay rate \cite{Minganti-18}.

We define the relaxation time of the system.
The expectation value of a local observable $\hat{O}$ at time $t$ and that for the steady state are denoted as $O(t)=\mathrm{Tr}[\hat{O}\hat{\rho}(t)]$ and $O_{\mathrm{ss}}=\mathrm{Tr}[\hat{O}\hat{\rho}_{\mathrm{ss}}]$, respectively.
First, we define $\tilde{\tau}(\hat{\rho}_{\mathrm{ini}},\hat{O})$ as the largest time $t$ that satisfies $|O(t)-O_{\mathrm{ss}}|\geq e^{-1}|O(0)-O_{\mathrm{ss}}|$, where we assume that $O(0) \neq O_{\mathrm{ss}}$ to ensure the finiteness of $\tilde{\tau}$.
We also define the maximal relaxation time $\tau$ by taking the supremum of $\tilde{\tau}(\hat{\rho}_{\mathrm{ini}},\hat{O})$ over all $\hat{\rho}_{\mathrm{ini}}$ and $\hat{O}$ under the condition $O(0) \neq O_{\mathrm{ss}}$.
If one takes the initial state $\hat{\rho}_{\mathrm{ini}}=\hat{\rho}_{\mathrm{ss}}+c_j\hat{\rho}^{\mathrm{R}}_{j}+c_j^*(\hat{\rho}^{\mathrm{R}}_{j})^{\dag}$ with arbitrary $j\:(\neq0)$, the relaxation time is given by $\tilde{\tau}(\hat{\rho}_{\mathrm{ini}},\hat{O})=\mathrm{Re} [\lambda_j]^{-1}$.
By maximizing $\tilde{\tau}$ over $\hat{\rho}_{\mathrm{ini}}$, it is natural to expect
\begin{equation}
\tau \: \overset{?}{=} \: \frac{1}{\Delta}.
\label{tau_Delta}
\end{equation}
If Eq.~(\ref{tau_Delta}) were true, the necessary and sufficient condition for the divergence of the maximal relaxation time would be closing of the Liouvillian gap.

We show that Eq.~(\ref{tau_Delta}) does not hold if the system exhibits the Liouvillian skin effect.
For simplicity, we consider a single-particle system in a one-dimensional space of length $L$.
Let $|x\rangle$ be the state in which the particle is located at position $x$.
We assume that the matrix element of the first right (left) eigenmode is exponentially localized near the right (left) boundary: $|\langle x|\hat{\rho}^{\mathrm{R}}_1|y\rangle|\sim e^{-(2L-x-y)/\xi}$, $|\langle x|\hat{\rho}^{\mathrm{L}}_1|y\rangle|\sim e^{-(x+y)/\xi}$, where $\xi$ is the localization length.
Then, the overlap between them is exponentially small: $(\hat{\rho}^{\mathrm{L}}_1|\hat{\rho}^{\mathrm{R}}_1)\sim e^{-O(L/\xi)}$.
Note that the numerator of Eq.~(\ref{rho_expansion_coefficients}) is maximized and takes a value of $O(1)$ when $\hat{\rho}_{\mathrm{ini}}$ is localized near the left boundary.
Thus, the maximal $|c_1|$ over all initial states is proportional to $e^{O(L/\xi)}$.
It is reasonable to state that the system has reached a steady state if $|O(t)-O_{\mathrm{ss}}| \ll \| \hat{O} \|_{\mathrm{op}}$ for any local observable $\hat{O}$, where $\| ... \|_{\mathrm{op}}$ denotes the operator norm.
From Eq.~(\ref{rho_expansion_t}), we have $|\sum_{j>0} c_j e^{\lambda_j t} \mathrm{Tr}[\hat{O}\hat{\rho}^{\mathrm{R}}_j]| \ll \| \hat{O} \|_{\mathrm{op}}$, where each $|\mathrm{Tr}[\hat{O}\hat{\rho}^{\mathrm{R}}_j]|$ is bounded by $\| \hat{O} \|_{\mathrm{op}}$.
Since in a later stage of relaxation the slowest mode ($j=1$) is expected to be dominant in the sum of the left-hand side, the condition for the system to reach its steady state is given by $|c_1|e^{-t \Delta} \ll 1$.
Thus, the maximal relaxation time $\tau$ is given by $|c_1|e^{-\tau \Delta}=e^{-1}$. 
From $|c_1| \sim e^{O(L/\xi)}$, we find
\begin{equation}
\tau \sim \frac{1}{\Delta} + \frac{L}{\xi \Delta}.
\label{tau_Delta_xi}
\end{equation}
This is the main result of this Letter.
In the absence of the skin effect, i.e., $\xi=L$, Eq.~(\ref{tau_Delta_xi}) reduces to Eq.~(\ref{tau_Delta}).
Equation (\ref{tau_Delta_xi}) implies that, if the Liouvillian gap and the localization length are independent of the system size, the maximal relaxation time is proportional to the system size.
Defining the relaxation velocity by $v_{\mathrm{R}}:=L/\tau$, Eq.~(\ref{tau_Delta_xi}) gives
\begin{equation}
v_{\mathrm{R}} \sim \xi \Delta
\label{vR_Delta_xi}
\end{equation}
for $L\to\infty$.
It should be noted that the derivation of Eq.~(\ref{tau_Delta_xi}) does not rely on the details of the Liouvillian superoperator, as long as its eigenmodes exhibit the skin effect.
While we here focus on the Lindblad master equation, the same argument can also be applied to other types of master equations, such as the Redfield equation.

We here note the relationship between the relaxation time $\tau$ defined above and the mixing time in Markov processes \cite{Levin, Diaconis-96, Kastoryano-12, Kastoryano-13, Vernier-20}.
The maximal distance to the steady state over initial states is given by $d(t) = \max_{\hat{\rho}_{\mathrm{ini}}} \| \hat{\rho}(t)-\hat{\rho}_{\mathrm{ss}} \|_{\mathrm{tr}}$.
The mixing time $\tau_{\mathrm{mix}}$ is then defined as the time for $d(t)$ to reach some small value $\epsilon$.
In general, $\tau_{\mathrm{mix}}$ provides an upper bound on the relaxation time of observables.
When $\hat{\rho}_{\mathrm{ss}}$ is localized, $\max_{\hat{\rho}_{\mathrm{ini}}} \tilde{\tau}(\hat{\rho}_{\mathrm{ini}},\hat{O})$ attains $\tau_{\mathrm{mix}}$ for local observables $\hat{O}$ with support in the localized region of $\hat{\rho}_{\mathrm{ss}}$, and thus the maximal relaxation time $\tau$ over all local observables coincides with $\tau_{\mathrm{mix}}$.
Therefore, Eq.~(\ref{tau_Delta_xi}) is also valid for the mixing time.

{\em Prototypical model.--}
Here, we present a prototypical model that shows the Liouvillian skin effect.
The Hamiltonian of the system is given by $\hat{H}=-J\sum_{l=1}^{L}(\hat{b}_{l+1}^{\dag} \hat{b}_{l}+\hat{b}_{l}^{\dag}\hat{b}_{l+1})$, where $\hat{b}^{\dag}_{l}$ and $\hat{b}_{l}$ are the creation and annihilation operators of a boson at site $l$, which satisfy $[\hat{b}_{l},\hat{b}_{m}^{\dag}]=\delta_{lm}$ and $[\hat{b}_{l},\hat{b}_{m}]=[\hat{b}_{l}^{\dag},\hat{b}_{m}^{\dag}]=0$, and $J$ represents the transfer amplitude.
We consider Lindblad operators $\hat{L}_{\mathrm{R},l}=\sqrt{\gamma_{\mathrm{R}}}\hat{b}_{l+1}^{\dag}\hat{b}_{l}$ and $\hat{L}_{\mathrm{L},l}=\sqrt{\gamma_{\mathrm{L}}}\hat{b}_{l-1}^{\dag}\hat{b}_{l}$ \cite{Eisler-11, Temme-12}, which describe stochastic hopping to the right and left neighboring sites with rates $\gamma_{\mathrm{R}}$ and $\gamma_{\mathrm{L}}$, respectively.
The index $\alpha$ in Eq.~(\ref{Liouvillian}) includes $\mathrm{R}$ or $\mathrm{L}$, and site index $l=1,2,...,L$.
We will discuss both cases of the open boundary condition (OBC) and the periodic boundary condition (PBC).
Given the state $|l\rangle$ in which the particle is located at site $l$, the set of vectors $\{|l\rangle\}_{l=1,.., L}$ forms an orthonormal basis of the Hilbert space.
The asymmetric stochastic hopping can be implemented with ultracold atoms in an optical lattice by a laser-assisted hopping with spontaneous emission \cite{Jaksch-03, Aidelsburger-13, Miyake-13, Dalibard-85, Pichler-10, Supplement}.
While such an experimental setup also gives rise to an on-site dephasing $\hat{L}_{\mathrm{d},l}=\sqrt{\gamma_{\mathrm{d}}}\hat{b}_{l}^{\dag}\hat{b}_{l}$, this additional dissipation does not affect the qualitative behaviors discussed below \cite{Supplement}.

{\em Relaxation time.--}
Suppose that the initial state is localized at a point with distance $d$ from the region in which the steady state is localized.
Since the total particle number $\hat{N}=\sum_{l=1}^L\hat{b}_{l}^{\dag}\hat{b}_{l}$ is conserved, the relaxation toward the steady state must be accompanied by the transport of particles.
In quantum dissipative systems with local interactions, there exists an upper bound on the speed at which information can propagate, i.e., the Lieb-Robinson bound \cite{Lieb-72, Poulin-10}.
Thus, at least it takes a time proportional to $d$ for the system to reach its steady state.
In other words, the maximal relaxation time diverges in the limit of infinite system size.

\begin{figure}
 \centering
 \includegraphics[width=0.45\textwidth]{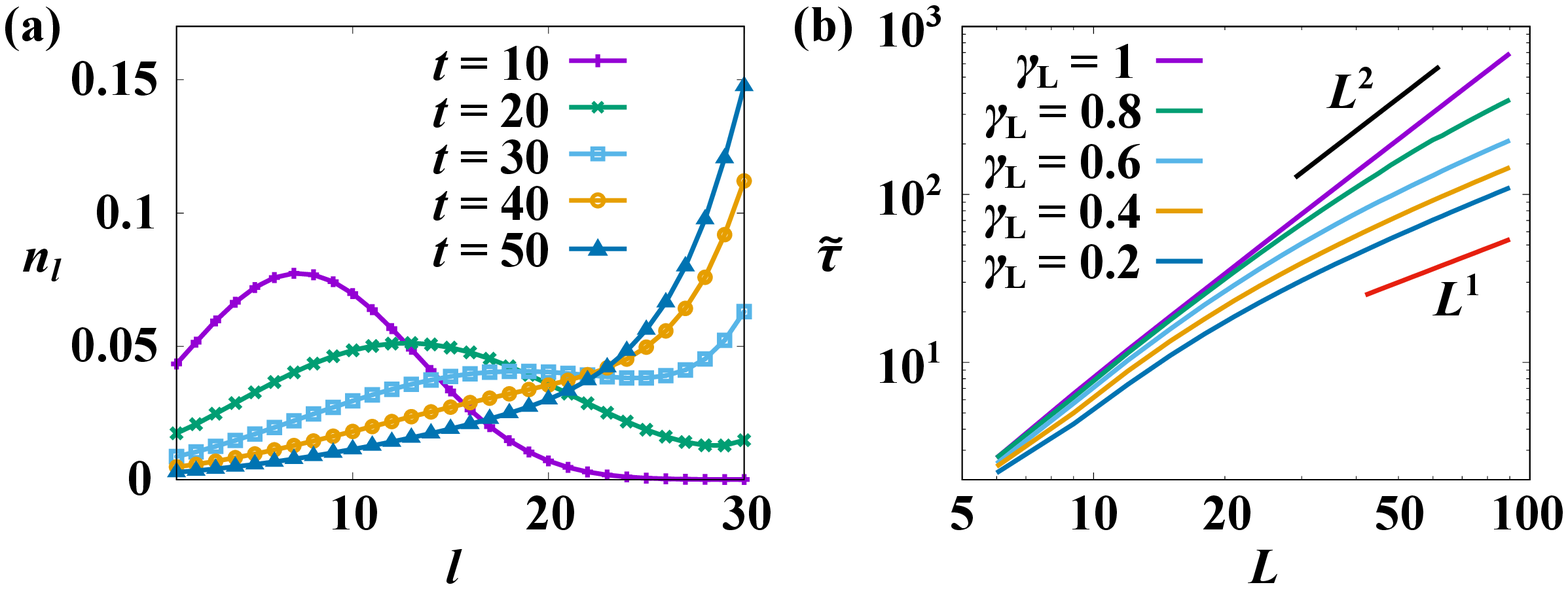}
 \caption{(a) Time evolution of the density profile $n_l$ with  hopping parameters $J=\gamma_{\mathrm{R}}=1$, $\gamma_{\mathrm{L}}=0.5$, and system size $L=30$.
 (b) Relaxation time $\tilde{\tau}$ with hopping parameters $J=\gamma_{\mathrm{R}}=1$, and $\gamma_{\mathrm{L}}=1$, $0.8$, $0.6$, $0.4$, $0.2$ from top to bottom.
 The abscissa and ordinate are shown in log scales.
 The two straight lines represent the $L^1$ and $L^2$ scalings.}
 \label{Fig-tau}
\end{figure}

Let us confirm this fact by numerically solving the master equation under the OBC.
Figure \ref{Fig-tau} (a) shows the time evolution of the density profile $n_l=\mathrm{Tr}[\hat{\rho} \hat{b}_{l}^{\dag}\hat{b}_{l}]=\langle l|\hat{\rho}|l\rangle$ from an initial state $\hat{\rho}_{\mathrm{ini}}=|1\rangle\langle1|$.
For $\gamma_{\mathrm{R}}>\gamma_{\mathrm{L}}$, the particle is transported from left to right, and accumulated at the right boundary.
Figure \ref{Fig-tau} (b) shows $\tilde{\tau}$ determined from the condition $n_{\mathrm{ss},L}-n_L(\tilde{\tau})=e^{-1}n_{\mathrm{ss},L}$, where $n_{\mathrm{ss},l}$ is the density profile of the steady state.
For $\gamma_{\mathrm{R}}=\gamma_{\mathrm{L}}$, $\tilde{\tau}$ is proportional to $L^2$, reflecting the diffusive relaxation to the uniform steady state.
In contrast, for $\gamma_{\mathrm{R}}>\gamma_{\mathrm{L}}$, $\tilde{\tau}$ is asymptotically proportional to $L$.
If the relation (\ref{tau_Delta}) were correct, the gap should always close.

{\em Liouvillian spectrum.--}
The operator space is spanned by $\{|l\rangle\langle m|\}_{l,m=1,...,L}$.
First, we consider the case of $J=0$,
where the action of $\mathcal{L}$ is closed in the diagonal subspace spanned by $\{|l \rangle\langle l|\}_{l=1,...,L}$ and in the off-diagonal subspace spanned by $\{|l\rangle\langle m|\}_{l,m=1,...,L;\: l \neq m}$ \cite{Supplement}.
If we interpret $|l\rangle\langle l|$ as the state in which a particle sits at site $l$, then $\mathcal{L}$ restricted to the diagonal subspace is equivalent to a non-Hermitian tight-binding Hamiltonian
\begin{equation}
\hat{\mathcal{H}}_{\mathrm{PBC}} = \sum_{l=1}^{L} ( \gamma_{\mathrm{R}} \hat{c}_{l+1}^{\dag} \hat{c}_l + \gamma_{\mathrm{L}} \hat{c}_{l}^{\dag} \hat{c}_{l+1} )
\label{HN_Hamiltonian_PBC}
\end{equation}
for the PBC, and
\begin{equation}
\hat{\mathcal{H}}_{\mathrm{OBC}} = \sum_{l=1}^{L-1} ( \gamma_{\mathrm{R}} \hat{c}_{l+1}^{\dag} \hat{c}_l + \gamma_{\mathrm{L}} \hat{c}_{l}^{\dag} \hat{c}_{l+1} )  + \gamma_{\mathrm{L}} \hat{c}_{1}^{\dag} \hat{c}_{1} +  \gamma_{\mathrm{R}} \hat{c}_{L}^{\dag} \hat{c}_{L}
\label{HN_Hamiltonian_OBC}
\end{equation}
for the OBC, where $\hat{c}_{l}^{\dag}$ and $\hat{c}_{l}$ are the creation and annihilation operators of the virtual particle, and we have omitted a constant energy shift $-\gamma_{\mathrm{R}}-\gamma_{\mathrm{L}}$.
Such a tight-binding model with asymmetric hopping is known as the Hatano-Nelson model \cite{Gong-18, Hatano-96, Hatano-97, Hatano-98}.

For the case of the PBC, a right eigenmode of $\hat{\mathcal{H}}_{\mathrm{PBC}}$ is given by a plane wave $\psi_{k,l}\propto e^{ikl}$ with $k=2\pi n/L \:(n=-L/2+1,...,L/2)$, and its eigenvalue reads
\begin{equation}
\lambda_k^{(\mathrm{PBC})} = \gamma_{\mathrm{R}} e^{- i k} + \gamma_{\mathrm{L}} e^{ i k} - \gamma_{\mathrm{R}} - \gamma_{\mathrm{L}},
\label{HN_lambda_PBC}
\end{equation}
where we have restored the constant shift $-\gamma_{\mathrm{R}}-\gamma_{\mathrm{L}}$.
From Eq.~(\ref{HN_lambda_PBC}), we have $\Delta\sim L^{-2}$.

\begin{figure}
 \centering
 \includegraphics[width=0.45\textwidth]{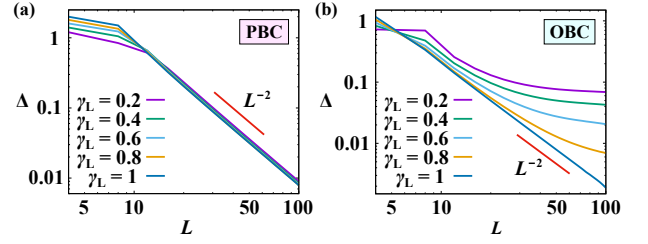}
 \caption{Liouvillian gap $\Delta$ as a function of the system size $L$ for the PBC (a) and the OBC (b) with $J=\gamma_{\mathrm{R}}=1$ and $\gamma_{\mathrm{L}}=0.2$, $0.4$, $0.6$, $0.8$, $1$.
 The abscissa and ordinate are shown in log scales.
 The straight solid lines represent the $L^{-2}$ scaling.}
 \label{Fig-gap}
\end{figure}

Next, we consider the eigenmodes of $\hat{\mathcal{H}}_{\mathrm{OBC}}$.
We introduce an imaginary gauge transformation $\hat{V}=\exp[-\ln r\sum_{l}l\hat{c}_l^{\dag}\hat{c}_l]$, which gives $\hat{V}^{-1}\hat{c}_l^{\dag}\hat{V}=r^{l}\hat{c}_l^{\dag}$ and $\hat{V}^{-1}\hat{c}_l\hat{V}=r^{-l}\hat{c}_l$ \cite{Hatano-96, Hatano-97, Hatano-98}.
When $r=\sqrt{\gamma_{\mathrm{L}}/\gamma_{\mathrm{R}}}$, the transformed Hamiltonian $\hat{\mathcal{H}}'_{\mathrm{OBC}}:=\hat{V}^{-1}\hat{\mathcal{H}}_{\mathrm{OBC}}\hat{V}$ is given by replacing $\gamma_{\mathrm{R}}$ and $\gamma_{\mathrm{L}}$ in the hopping term $\gamma_{\mathrm{R}}\hat{c}_{l+1}^{\dag}\hat{c}_l+\gamma_{\mathrm{L}}\hat{c}_{l}^{\dag}\hat{c}_{l+1}$ in Eq.~(\ref{HN_Hamiltonian_OBC}) with $\sqrt{\gamma_{\mathrm{R}}\gamma_{\mathrm{L}}}$.
Note that the eigenspectrum of $\hat{\mathcal{H}}_{\mathrm{OBC}}$ is real because $\hat{\mathcal{H}}'_{\mathrm{OBC}}$ is Hermitian.
Let $\psi_l$ and $\psi'_l$ be a right eigenmode of $\hat{\mathcal{H}}_{\mathrm{OBC}}$ and that of $\hat{\mathcal{H}}'_{\mathrm{OBC}}$ sharing the same eigenvalue.
The right  eigenmodes of $\hat{\mathcal{H}}'_{\mathrm{OBC}}$ include a bound state $\psi'_{\mathrm{BS},l}\propto(\gamma_{\mathrm{R}}/\gamma_{\mathrm{L}})^{l/2}$ and $(L-1)$ plane-wave states $\psi'_{k,l}=c_1e^{ikl}+c_2e^{-ikl}$ with $k=n\pi/L \:(n=1,...,L-1)$ \cite{Supplement}.
The bound state corresponds to the steady state.
The eigenvalues for the plane-wave states are given by
\begin{equation}
\lambda_k^{(\mathrm{OBC})} = 2 \sqrt{\gamma_{\mathrm{R}}\gamma_{\mathrm{L}}} \cos k - \gamma_{\mathrm{R}} - \gamma_{\mathrm{L}}.
\label{HN_lambda_OBC}
\end{equation}
The eigenmode of the original Hamiltonian $\hat{\mathcal{H}}_{\mathrm{OBC}}$ is given by $\psi_l=(\gamma_{\mathrm{R}}/\gamma_{\mathrm{L}})^{l/2}\psi'_l$.
When $\gamma_{\mathrm{R}}>\gamma_{\mathrm{L}}$, all right eigenmodes are exponentially localized near the right boundary.
From Eq.~(\ref{HN_lambda_OBC}), the gap is given by
\begin{equation}
\Delta = \gamma_{\mathrm{R}} + \gamma_{\mathrm{L}} - 2 \sqrt{\gamma_{\mathrm{R}}\gamma_{\mathrm{L}}}
\end{equation}
in the $L \to \infty$ limit.
Note that for $\gamma_{\mathrm{R}}=\gamma_{\mathrm{L}}$, the gap closes as $L^{-2}$.
Thus, in sharp contrast to the case of the PBC, the Liouvillian spectrum under the OBC has a nonvanishing gap for $\gamma_{\mathrm{R}}\neq\gamma_{\mathrm{L}}$.
Such an extreme sensitivity of the eigenspectrum to the boundary conditions is a special character of quantum dissipative systems driven out of equilibrium, and reminiscent of a similar effect seen in the non-Hermitian systems \cite{Lee-16, Gong-18, Yao-18, Kunst-18, Thomale-18, Song-19, Borgnia-20, Okuma-20, Okugawa-20, Kawabata-20}.
The left eigenmodes of $\hat{\mathcal{H}}_{\mathrm{OBC}}$ can be obtained by exchanging $\gamma_{\mathrm{R}}$ and $\gamma_{\mathrm{L}}$ in the right eigenmodes.
Thus, for $\gamma_{\mathrm{R}}>\gamma_{\mathrm{L}}$, the left eigenmodes are localized near the left boundary.

The eigenvalues given by Eqs.~(\ref{HN_lambda_PBC}) and (\ref{HN_lambda_OBC}) are those of $\mathcal{L}$ that are restricted to the diagonal subspace.
In addition to them, there are $(L^2-L)$ eigenvalues belonging to the off-diagonal subspace.
For the PBC, it is given by $\lambda=-\gamma_{\mathrm{R}}-\gamma_{\mathrm{L}}$.
For the OBC, there are four eigenvalues $\lambda=-\gamma_{\mathrm{R}}-\gamma_{\mathrm{L}}$, $-\gamma_{\mathrm{R}}-\gamma_{\mathrm{L}}/2$, $-\gamma_{\mathrm{R}}/2-\gamma_{\mathrm{L}}$, and $-\gamma_{\mathrm{R}}/2-\gamma_{\mathrm{L}}/2$, whose degeneracies are $(L-2)(L-3)$, $2(L-2)$, $2(L-2)$, and $2$, respectively.

For $J\neq0$, the eigenvalue problem of the Liouvillian cannot be solved exactly, and we study its eigenspectrum using numerical diagonalization.
Figure \ref{Fig-gap} shows $\Delta$ as a function of $L$.
For the PBC case (a), $\Delta$ vanishes as $L^{-2}$ for arbitrary hopping parameters.
For the OBC case (b), while $\Delta \sim L^{-2}$ for $\gamma_{\mathrm{R}}=\gamma_{\mathrm{L}}$, $\Delta$ approaches a nonzero value for $\gamma_{\mathrm{R}}\neq\gamma_{\mathrm{L}}$ \cite{Znidaric-15}.

\begin{figure}
 \centering
 \includegraphics[width=0.45\textwidth]{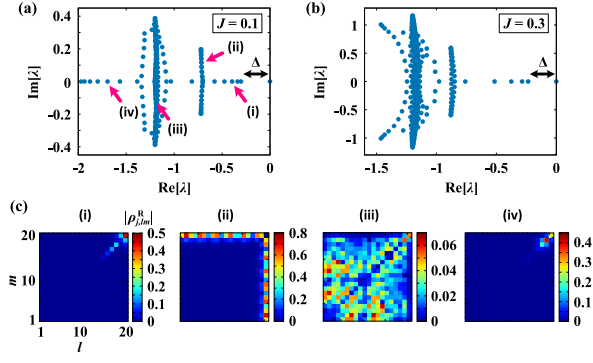}
 \caption{(a), (b) Liouvillian eigenspectrum for the OBC with $\gamma_{\mathrm{R}}=1$, $\gamma_{\mathrm{L}}=0.2$, and $J=0.1$ (a) and $J=0.3$ (b).
 The system size is $L=20$.
 (c) Color plots of $|\rho^{\mathrm{R}}_{j,lm}|$ corresponding to the eigenvalues indicated by arrows $\rm(\hspace{.1em}i\hspace{.1em})$ -- $\rm(i\hspace{-.08em}v\hspace{-.06em})$ in (a).}
 \label{Fig-spec-eigvec}
\end{figure}

Next, we show that the eigenmodes exhibit the Liouvillian skin effect for nonzero $J$.
Figures \ref{Fig-spec-eigvec} (a) and (b) show $\{ \lambda_{j} \}_{j=0,...,L^2-1}$ for the OBC.
The eigenspectrum is composed of two groups of eigenvalues with and without imaginary parts.
The real spectrum stems from the eigenvalues given by Eq.~(\ref{HN_lambda_OBC}) at $J=0$.
One can clearly see two clusters of complex eigenvalues around $\mathrm{Re}[\lambda]=-\gamma_{\mathrm{R}}-\gamma_{\mathrm{L}}=-1.2$ and $\mathrm{Re}[\lambda]=-\gamma_{\mathrm{R}}/2-\gamma_{\mathrm{L}}=-0.7$, which originate from the highly degenerated eigenvalues at $J=0$.
Figure \ref{Fig-spec-eigvec} (c) shows the modulus of $\rho^{\mathrm{R}}_{j,lm}=\langle l|\hat{\rho}^{\mathrm{R}}_j|m\rangle$ for the eigenvalues indicated by arrows $\rm(\hspace{.1em}i\hspace{.1em})$ -- $\rm(i\hspace{-.08em}v\hspace{-.06em})$ in Fig.~\ref{Fig-spec-eigvec} (a).
For the real eigenvalues such as $\rm(\hspace{.1em}i\hspace{.1em})$ and $\rm(i\hspace{-.08em}v\hspace{-.06em})$, $\rho^{\mathrm{R}}_{j,lm}$ is localized exponentially near the right boundary.
In contrast, for the complex eigenvalues such as $\rm(i\hspace{-.05em}i)$ and $\rm(i\hspace{-.05em}i\hspace{-.05em}i)$, $\rho^{\mathrm{R}}_{j,lm}$ is delocalized for one or both of $l$ and $m$.
The partial skin effect observed in $\rm(i\hspace{-.05em}i)$ can be understood from the fact that $\hat{\rho}^{\mathrm{R}}_{j}$ is a superposition of $|l\rangle\langle m|$ where either $l$ or $m$ belongs to the right boundary,  which is the eigenmode with $\lambda=-\gamma_{\mathrm{R}}/2-\gamma_{\mathrm{L}}$ at $J=0$.
The eigenmodes with the complex eigenvalues are irrelevant to the slowing down of relaxation because they do not exhibit the Liouvillian skin effect.

\begin{figure}
 \centering
 \includegraphics[width=0.45\textwidth]{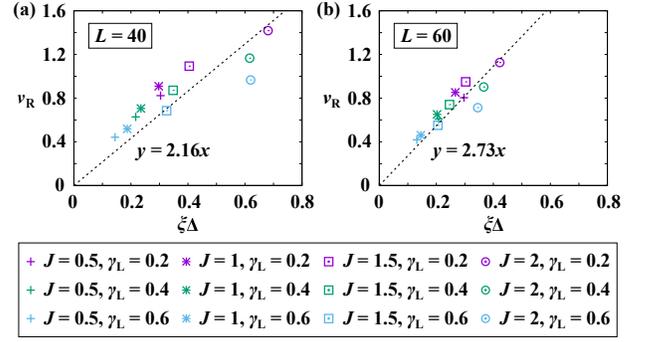}
 \caption{Plots of $v_{\mathrm{R}}$ versus $\xi\Delta$ with $\gamma_{\mathrm{R}}=1$, and $J=0.5$ (cross), $1$ (star), $1.5$ (square), $2$ (circle), and $\gamma_{\mathrm{L}}=0.2$ (purple), $0.4$ (green), $0.6$ (blue).
 The system sizes are (a) $L=40$ and (b) $L=60$.
 The dashed lines represent the least squares fitting by $y=ax$.
 The relative standard error of $a$ is $6.5\%$ for (a) and $4.0\%$ for (b).}
 \label{Fig-gap-xi-vR}
\end{figure}

Let us confirm Eq.~(\ref{vR_Delta_xi}) for the prototypical model.
We assume that $\xi$ is of the same order of magnitude as the localization length of the steady state.
By using the density profile $n_{\mathrm{ss},l}$ of the steady state, $\xi$ is estimated to be $\xi=1/n_{\mathrm{ss},L}$.
Figure \ref{Fig-gap-xi-vR} shows $v_{\mathrm{R}}$ and $\xi\Delta$ for several parameters.
As the system size increases, the plots of $v_{\mathrm{R}}$ versus $\xi\Delta$ converge to points on a single line $v_{\mathrm{R}}=a\xi\Delta$ with $a\simeq2.7$.

To estimate the relaxation time in Figs.~\ref{Fig-tau} and \ref{Fig-gap-xi-vR}, we have focused on the diagonal elements $\rho_{ll}=\langle l|\hat{\rho}|l\rangle$ of the density matrix.
It is also intriguing to consider the relaxation of the off-diagonal elements $\rho_{lm}=\langle l|\hat{\rho}|m\rangle\:(l\neq m)$, because they provide a measure of quantum coherence.
For $J=0$, since $\hat{\rho}_1^{\mathrm{R}}$ does not contain the off-diagonal elements, $\rho_{lm}$ decays at a constant rate and Eq.~(\ref{tau_Delta_xi}) does not hold.
In contrast, for $J\neq0$, the relaxation time of $\rho_{lm}$ diverges with the system size \cite{Supplement} since the eigenmodes exhibiting the skin effect have nonzero off-diagonal elements (see $\rm(\hspace{.1em}i\hspace{.1em})$ in Fig.~\ref{Fig-spec-eigvec}).

{\em Generalizations.--}
The prototypical model can be generalized to an $N\:(<L)\:$-particle system with the hard-core condition $(\hat{b}_l^{\dag})^2=0$.
For $J=0$, the model is identical to a classical Markov process known as the asymmetric simple exclusion process (ASEP) \cite{Derrida-98}.
For the OBC, it has been proven that the transition matrix of the ASEP has a nonzero spectral gap in the thermodynamic limit \cite{Sandow-94}.
The density profile $n_l$ of the steady state is given by $n_{L-N-l}\propto e^{-l/\xi}$ and $n_{L-N+l}\propto1-e^{-l/\xi}$ when $\gamma_{\mathrm{R}}>\gamma_{\mathrm{L}}$.
The existence of such an exponential tail in the density profile suggests that the overlap between the right and left eigenmodes should be exponentially small.
Thus, Eq.~(\ref{tau_Delta_xi}) should hold for $J=0$.
It merits further study to investigate the validity of Eq.~(\ref{tau_Delta_xi}) for the many-body case with nonzero $J$.

The non-Hermitian skin effects in higher-dimensional systems are currently under active investigation \cite{Okuma-20, Okugawa-20, Kawabata-20}.
Our results can also be extended to higher-dimensional cases.
For example, a possible manifestation of the skin effect in a rectangular-shaped two-dimensional system is the localization of the right eigenmodes at one of the four edges of the boundary. 
If the left eigenmodes are localized at the opposite side, Eq.~(\ref{tau_Delta_xi}) holds by replacing the one-dimensional system size $L$ by the length perpendicular to the edges at which the eigenmodes are localized.
Understanding what factors determine the location of the eigenmode localization in higher-dimensional systems merits further study.

Finally, we remark on the non-Markovian effect in the dynamics of open quantum systems.
As in the Markovian cases, one can also perform the spectral decomposition of the propagator corresponding to the temporally nonlocal master equation \cite{Chruscinski-10}.
A major difference from the Markovian case is that the eigenmodes and the spectrum explicitly depend on time. 
If these time-dependent eigenmodes are always localized near the boundary, we can expect the slowing down of relaxation processes without gap closing.

{\em Conclusions.--}
We have shown that the longest relaxation time of a quantum dissipative system depends not only on the Liouvillian gap, but also on the localization length of the eigenmodes.
This behavior is a nonequilibrium effect unique to dissipative systems driven far from thermal equilibrium.
In fact, it can be shown that, if the Liouvillian with a spatially uniform Hamiltonian satisfies the detailed balance condition, the right and left eigenmodes cannot be localized near opposite boundaries of the system \cite{Supplement}.
It is worthwhile to study other mechanisms that lead to a small overlap between the right and left eigenmodes \cite{Diaconis-96, Kastoryano-12, Kastoryano-13, Vernier-20, Mori-20}.

\begin{acknowledgments}
{\em Acknowledgments.--}
The authors thank T. Mori for helpful discussions.
This work was supported by KAKENHI Grant Numbers JP19J00525 and JP18H01145, and a Grant-in-Aid for Scientific Research on Innovative Areas (KAKENHI Grant Number JP15H05855) from Japan Society for the Promotion of Science (JSPS).
M. N. was supported by KAKENHI Grant Number JP20K14383.
R. H. was supported by JSPS through Program for Leading Graduate Schools (ALPS) and JSPS fellowship (KAKENHI Grant Number JP17J03189). 
\end{acknowledgments}

\clearpage
\widetext
\setcounter{equation}{0}
\def\theequation{S\arabic{equation}}
\setcounter{figure}{0}
\def\thefigure{S\arabic{figure}}
\appendix

\section{SUPPLEMENTAL MATERIAL}

\subsection{S1. \ Eigenmodes of a prototypical model in the absence of the coherent hopping}

In this section, we calculate the eigenmodes and the corresponding eigenvalues of a prototypical model with $J=0$, where a particle moves on a lattice solely through stochastic hopping.
For the periodic boundary condition, the action of the Liouvillian is given by
\begin{equation}
\mathcal{L}(|l \rangle \langle l|) = \gamma_{\mathrm{R}} |l+1 \rangle \langle l+1| + \gamma_{\mathrm{L}} |l-1 \rangle \langle l-1| - (\gamma_{\mathrm{R}}+\gamma_{\mathrm{L}}) |l \rangle \langle l|,
\label{Supplement_L_diagonal}
\end{equation}
for a diagonal basis and
\begin{equation}
\mathcal{L}(|l \rangle \langle m|) = - (\gamma_{\mathrm{R}}+\gamma_{\mathrm{L}}) |l \rangle \langle m|, \: (l \neq m),
\label{Supplement_L_off_diagonal}
\end{equation}
for an off-diagonal basis.
From Eqs.~(\ref{Supplement_L_diagonal}) and (\ref{Supplement_L_off_diagonal}), we find that the action of the Liouvillian is decoupled into the diagonal and off-diagonal subspaces.
The eigenvalue associated with the off-diagonal subspace is trivially given by $-\gamma_{\mathrm{R}}-\gamma_{\mathrm{L}}$.
We formally identify a diagonal basis $|l \rangle \langle l|$ with a state $| l )$ in which a virtual particle resides at site $l$.
Then, Eq.~(\ref{Supplement_L_diagonal}) can be rewritten as
\begin{equation}
\mathcal{L} | l ) = \gamma_{\mathrm{R}} | l+1 ) + \gamma_{\mathrm{L}} | l-1 ) - (\gamma_{\mathrm{R}}+\gamma_{\mathrm{L}}) | l ).
\label{Supplement_L_diagonal_2}
\end{equation}
By introducing the creation and annihilation operators $\hat{c}_{l}^{\dag}$ and $\hat{c}_{l}$ of the virtual particle, the action (\ref{Supplement_L_diagonal_2}) of the Liouvillian is identical to that of a non-Hermitian tight-binding Hamiltonian
\begin{eqnarray}
\hat{\mathcal{H}}_{\mathrm{PBC}} &=& \sum_{l=1}^{L} \bigl[ \gamma_{\mathrm{R}} \hat{c}_{l+1}^{\dag} \hat{c}_l + \gamma_{\mathrm{L}} \hat{c}_{l}^{\dag} \hat{c}_{l+1} - ( \gamma_{\mathrm{R}} + \gamma_{\mathrm{L}} ) \hat{c}_{l}^{\dag} \hat{c}_l \bigr].
\label{Supplement_HN_Hamiltonian_PBC}
\end{eqnarray}
For the open-boundary case, we have $\mathcal{L}(|1 \rangle \langle 1|)=\gamma_{\mathrm{R}}|2 \rangle \langle 2|-\gamma_{\mathrm{R}}|1 \rangle \langle 1|$ for the left boundary and $\mathcal{L}(|L \rangle \langle L|)=\gamma_{\mathrm{L}}|L-1 \rangle \langle L-1|-\gamma_{\mathrm{L}}|L \rangle \langle L|$ for the right boundary.
The same mapping from $|l \rangle \langle l|$ to $| l )$ yields the following Hamiltonian:
\begin{equation}
\hat{\mathcal{H}}_{\mathrm{OBC}} = \sum_{l=1}^{L-1} ( \gamma_{\mathrm{R}} \hat{c}_{l+1}^{\dag} \hat{c}_l + \gamma_{\mathrm{L}} \hat{c}_{l}^{\dag} \hat{c}_{l+1} ) - ( \gamma_{\mathrm{R}} + \gamma_{\mathrm{L}} ) \sum_{l=1}^L \hat{c}_{l}^{\dag} \hat{c}_l + \gamma_{\mathrm{L}} \hat{c}_{1}^{\dag} \hat{c}_{1} +  \gamma_{\mathrm{R}} \hat{c}_{L}^{\dag} \hat{c}_{L}.
\label{Supplement_HN_Hamiltonian_OBC}
\end{equation}
If we ignore the constant energy shift $-\gamma_{\mathrm{R}}-\gamma_{\mathrm{L}}$, Eqs.~(\ref{Supplement_HN_Hamiltonian_PBC}) and (\ref{Supplement_HN_Hamiltonian_OBC}) reduce to Eqs.~(10) and (11) in the main text, respectively.
It is worth noting that the matrix representation of $\hat{\mathcal{H}}$ is identical to the transition matrix of an asymmetric random walk process with hopping rates $\gamma_{\mathrm{R}}$ and $\gamma_{\mathrm{L}}$.

\subsubsection{A. \ Periodic boundary condition}

For the case of the periodic boundary condition (PBC), the eigenvalue equation reads
\begin{equation}
\gamma_{\mathrm{R}} \psi_{l-1} + \gamma_{\mathrm{L}} \psi_{l+1} - ( \gamma_{\mathrm{R}} + \gamma_{\mathrm{L}} ) \psi_l = \lambda \psi_l.
\label{Supplement_eigen_eq_PBC}
\end{equation}
The solution of Eq.~(\ref{Supplement_eigen_eq_PBC}) is given by a plane wave
\begin{equation}
\psi_{k,l} = e^{ikl},
\label{Supplement_eigvec_PBC}
\end{equation}
where $k = 2\pi n/L \ (n=-L/2+1,...,L/2)$.
The eigenvalue associated with Eq.~(\ref{Supplement_eigvec_PBC}) is calculated as
\begin{equation}
\lambda_k^{(\mathrm{PBC})} = \gamma_{\mathrm{R}} e^{- i k} + \gamma_{\mathrm{L}} e^{ i k} - \gamma_{\mathrm{R}} - \gamma_{\mathrm{L}}.
\label{Supplement_HN_lambda_PBC}
\end{equation}
The eigenmode with $k=0$ is the zero mode of $\hat{\mathcal{H}}_{\mathrm{PBC}}$, which corresponds to the steady state of the master equation.
Since the eigenvalue with the smallest modulus of the real part is given by $\lambda_{k=\pm2\pi/L}^{(\mathrm{PBC})}$, the spectral gap $\Delta$ is given by
\begin{equation}
\Delta = (\gamma_{\mathrm{R}}+\gamma_{\mathrm{L}}) \left[ 1 - \cos \left( \frac{2\pi}{L} \right) \right] \sim L^{-2}.
\end{equation}

\begin{figure}
 \centering
 \includegraphics[width=0.6\textwidth]{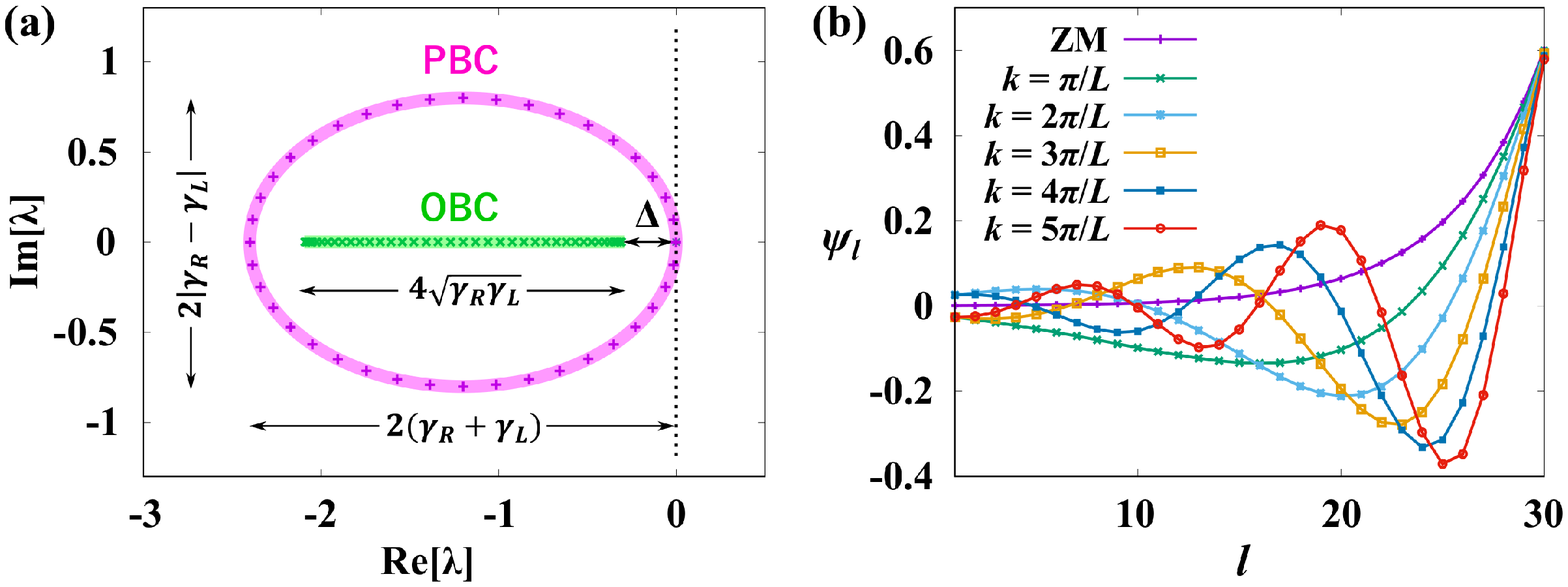}
 \caption{(a) Eigenspectrum of the Hatano-Nelson Hamiltonian for the PBC and OBC with hopping parameters $\gamma_{\mathrm{R}}=1$ and $\gamma_{\mathrm{L}}=0.2$, and the system size $L=40$. 
 The thick lines represent the eigenspectrum in the limit $L \to \infty$.
 (b) Eigenmodes of $\hat{\mathcal{H}}_{\mathrm{OBC}}$ with hopping parameters $\gamma_{\mathrm{R}}=1$ and $\gamma_{\mathrm{L}}=0.8$, and the system size $L=30$ for several different wave numbers $k$ of the eigenmode $\psi_{k,l}$, which is obtained from the plane-wave solution (\ref{Supplement_eigvec_OBC}).
 The zero mode (ZM) is given by $\psi_l \propto (\gamma_{\mathrm{R}}/\gamma_{\mathrm{L}})^l$.
 All eigenmodes including the zero mode are normalized as $\sum_l |\psi_l|^2=1$.}
 \label{Supplement-Fig-spectrum-HN}
\end{figure}

The eigenspectrum (\ref{Supplement_HN_lambda_PBC}) is shown in Fig.~\ref{Supplement-Fig-spectrum-HN} (a) in purple.
It forms an ellipse tangential to the imaginary axis (dotted line).
The diameters of the ellipse in the real and imaginary directions are given by $2(\gamma_{\mathrm{R}}+\gamma_{\mathrm{L}})$ and $2|\gamma_{\mathrm{R}}-\gamma_{\mathrm{L}}|$, respectively.
When $\gamma_{\mathrm{R}}>\gamma_{\mathrm{L}}$, as $k$ increases from $-\pi$ to $\pi$, the eigenvalue circumnavigates the ellipse clockwise.
For $\gamma_{\mathrm{R}}=\gamma_{\mathrm{L}}$, the ellipse collapses onto the real axis.

\subsubsection{B. \ Open boundary condition}

Let us consider the case of the open boundary condition (OBC).
We introduce an imaginary gauge transformation
\begin{equation}
\hat{V} = \exp\biggl[-\ln r \sum_{l=1}^L l \hat{c}_l^{\dag}\hat{c}_l \biggr],
\end{equation}
which transforms the creation and annihilation operators as
\begin{equation}
\hat{V}^{-1} \hat{c}_l^{\dag} \hat{V} = r^{l} \hat{c}_l^{\dag}, \:\:\: \hat{V}^{-1} \hat{c}_l \hat{V} = r^{-l} \hat{c}_l.
\label{Supplement_imag_gauge_trans}
\end{equation}
If we choose $r=\sqrt{\gamma_{\mathrm{L}}/\gamma_{\mathrm{R}}}$, the non-Hermitian Hamiltonian (\ref{Supplement_HN_Hamiltonian_OBC}) can be transformed to a Hermitian Hamiltonian
\begin{equation}
\hat{\mathcal{H}}'_{\mathrm{OBC}}  = \hat{V}^{-1} \hat{\mathcal{H}}_{\mathrm{OBC}} \hat{V} = \tilde{\gamma} \sum_{l=1}^{L-1} ( \hat{c}_{l+1}^{\dag} \hat{c}_l + \hat{c}_{l}^{\dag} \hat{c}_{l+1} ) - ( \gamma_{\mathrm{R}} + \gamma_{\mathrm{L}} ) \sum_{l=1}^L \hat{c}_{l}^{\dag} \hat{c}_l + \gamma_{\mathrm{L}} \hat{c}_{1}^{\dag} \hat{c}_{1} +  \gamma_{\mathrm{R}} \hat{c}_{L}^{\dag} \hat{c}_{L},
\label{Supplement_HN_Hamiltonian_OBC'}
\end{equation}
where $\tilde{\gamma} \equiv \sqrt{\gamma_{\mathrm{R}}\gamma_{\mathrm{L}}}$.
Let $\psi'_l$ be an eigenmode of $\hat{\mathcal{H}}'_{\mathrm{OBC}}$.
The eigenvalue equation for $\hat{\mathcal{H}}'_{\mathrm{OBC}}$ is given by
\begin{eqnarray}
\left\{
\begin{array}{l}
\tilde{\gamma} \psi'_2 - \gamma_{\mathrm{R}} \psi'_1 = \lambda \psi'_1, \\
\tilde{\gamma} (\psi'_{l-1}+\psi'_{l+1}) - (\gamma_{\mathrm{R}} + \gamma_{\mathrm{L}}) \psi'_l = \lambda \psi'_l \quad (1 < l < L), \\
\tilde{\gamma} \psi'_{L-1} - \gamma_{\mathrm{L}} \psi'_L = \lambda \psi'_L.
\end{array}
\right.
\label{Supplement_eigen_eq_OBC}
\end{eqnarray}

First, one can confirm that Eq.~(\ref{Supplement_eigen_eq_OBC}) has a bound state with eigenvalue $\lambda=0$ and eigenfunction
\begin{equation}
\psi'_{\mathrm{BS},l} = \left( \frac{\gamma_{\mathrm{R}}}{\gamma_{\mathrm{L}}} \right)^{l/2},
\label{Supplement_psi_BS}
\end{equation}
where we omit a normalization constant.
Note that the potential terms at the boundaries $\gamma_{\mathrm{L}} \hat{c}_{1}^{\dag} \hat{c}_{1}$ and $\gamma_{\mathrm{R}} \hat{c}_{L}^{\dag} \hat{c}_{L}$ in Eq.~(\ref{Supplement_HN_Hamiltonian_OBC'}) ensure the existence of the zero mode (\ref{Supplement_psi_BS}), which corresponds to the steady state of the master equation.
We next assume that Eq.~(\ref{Supplement_eigen_eq_OBC}) has a plane-wave solution given by
\begin{equation}
\psi'_{k,l} = c_1 e^{ikl} + c_2 e^{-ikl}.
\label{psi_plane_wave}
\end{equation}
Substituting Eq.~(\ref{psi_plane_wave}) into (\ref{Supplement_eigen_eq_OBC}), we obtain $(L-1)$ solutions
\begin{eqnarray}
\psi'_{k,l} = ( \gamma_{\mathrm{L}} - 2 \tilde{\gamma} \cos k ) \sin k(l-1) + ( \gamma_{\mathrm{R}} - 2 \tilde{\gamma} \cos k ) \sin k(L-l) + \tilde{\gamma} \sin k(l-2) + \tilde{\gamma} \sin k(L-1-l)
\label{Supplement_eigvec_OBC}
\end{eqnarray}
with $k=n\pi/L \ (n=1,...,L-1)$.
The eigenvalue associated with Eq.~(\ref{Supplement_eigvec_OBC}) is calculated as
\begin{equation}
\lambda_k = - \gamma_{\mathrm{R}} - \gamma_{\mathrm{L}} + 2 \tilde{\gamma} \cos k.
\label{Supplement_HN_lambda_OBC}
\end{equation}
The spectral gap $\Delta$ is given by
\begin{equation}
\Delta = \gamma_{\mathrm{R}} + \gamma_{\mathrm{L}} - 2 \tilde{\gamma}
\end{equation}
in the limit of $L \to \infty$.
For $\gamma_{\mathrm{R}}=\gamma_{\mathrm{L}}$, the gap closes as $\Delta \sim L^{-2}$.
From Eq.~(\ref{Supplement_imag_gauge_trans}), the eigenmodes of the original Hamiltonian $\hat{\mathcal{H}}_{\mathrm{OBC}}$ and the transformed one $\hat{\mathcal{H}}'_{\mathrm{OBC}}$ are related to each other by $\psi_l = ( \gamma_{\mathrm{R}}/\gamma_{\mathrm{L}} )^{l/2} \psi'_l$.
Defining the localization length $\xi$ via $|\psi_{k,l}|^2 \sim e^{-(L-l)/\xi}$ for $\gamma_{\mathrm{R}}>\gamma_{\mathrm{L}}$, we have $\xi = |\ln (\gamma_{\mathrm{R}}/\gamma_{\mathrm{L}})|^{-1}$.
We note that $\xi \propto \delta \gamma^{-1}$ and $\Delta \propto \delta \gamma$ for $\delta \gamma \equiv |\gamma_{\mathrm{R}}-\gamma_{\mathrm{L}}| \ll \gamma_{\mathrm{R}}$.

The eigenspectrum (\ref{Supplement_HN_lambda_OBC}) is shown in Fig.~\ref{Supplement-Fig-spectrum-HN} (a) with green symbols on the real axis.
It extends over the width $4\sqrt{\gamma_{\mathrm{R}}\gamma_{\mathrm{L}}}$ around the center point $\lambda=-\gamma_{\mathrm{R}}-\gamma_{\mathrm{L}}$.
Note that for $\gamma_{\mathrm{R}}=\gamma_{\mathrm{L}} \equiv \gamma$, eigenvalues for the PBC and OBC collapse onto a single line of length $4\gamma$ on the real axis.
Figure \ref{Supplement-Fig-spectrum-HN} (b) shows the eigenmodes of $\hat{\mathcal{H}}_{\mathrm{OBC}}$.
When $\gamma_{\mathrm{R}}>\gamma_{\mathrm{L}}$, all eigenmodes are localized near the right boundary ($l=L$).
The localization length is estimated to be $\xi = |\ln (1/0.8)|^{-1} \simeq 4.48$.

\subsection{S2. \ Relaxation of the off-diagonal elements of the density matrix}

In the main text, we have focused on the relaxation of the diagonal elements of the density matrix $\rho_{ll} = \langle l | \hat{\rho} | l \rangle$.
The relaxation time associated with the diagonal elements diverges linearly with increasing the system size irrespective of the value of the transfer amplitude $J$.
In contrast, the behavior of the relaxation time associated with the off-diagonal elements $\rho_{lm} = \langle l | \hat{\rho} | m \rangle \:(l \neq m)$ depends significantly on the presence or absence of $J$.
When $J=0$, the time evolution of the off-diagonal elements is given by
\begin{eqnarray}
\frac{d\rho_{lm}}{dt} = \left\{ \begin{array}{ll}
-(\gamma_{\mathrm{R}}/2+\gamma_{\mathrm{L}}/2) \rho_{lm}, &(l=1, \:m=L \: \mathrm{or} \: l=L, \:m=1); \\
-(\gamma_{\mathrm{R}}/2+\gamma_{\mathrm{L}}) \rho_{lm}, &(l=L \: \mathrm{or} \:  m=L); \\
-(\gamma_{\mathrm{R}}+\gamma_{\mathrm{L}}/2) \rho_{lm}, &(l=1 \: \mathrm{or} \:  m=1); \\
-(\gamma_{\mathrm{R}}+\gamma_{\mathrm{L}}) \rho_{lm}, &(\mathrm{otherwise}). \\
\end{array} \right.
\end{eqnarray}
Thus, the relaxation time is independent of the system size.

\begin{figure}
 \centering
 \includegraphics[width=0.55\textwidth]{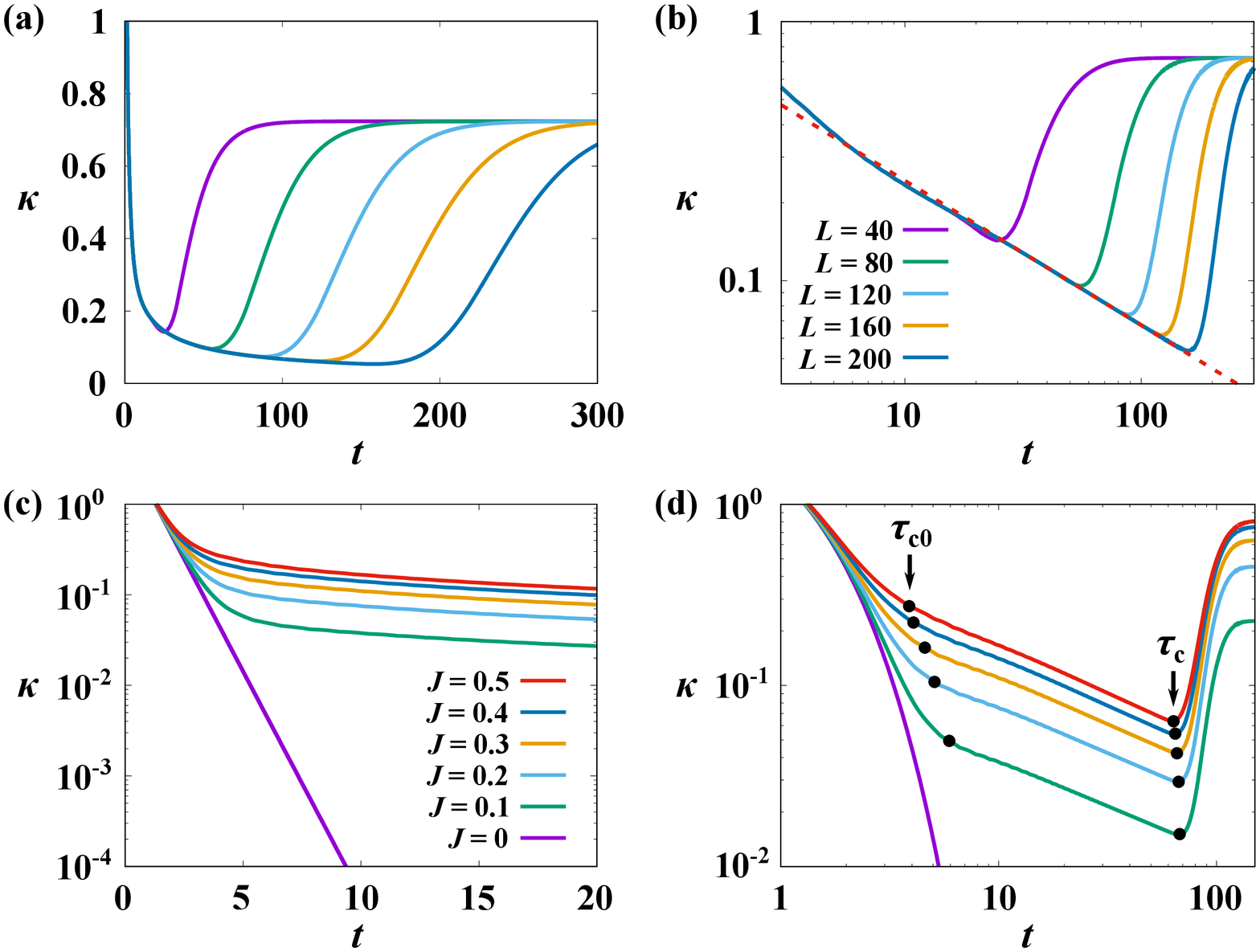}
 \caption{(a) Time evolution of $\kappa$ with hopping parameters $J=\gamma_{\mathrm{R}}=1$ and $\gamma_{\mathrm{L}}=0.2$, and system size $L=40$, $80$, $120$, $160$, $200$ from top to bottom.
 The standard deviation of the initial state is $\sigma = 2$.
 Figure (b) shows the same data as (a) in log scales, and the dashed line shows $\kappa \propto t^{-0.56}$.
 (c) Time evolution of $\kappa$ on a semi-log plot with the system size $L=80$ and hopping parameters $\gamma_{\mathrm{R}}=1$ and $\gamma_{\mathrm{L}}=0.2$ for $J=0$, $0.1$, $0.2$, $0.3$, $0.4$, $0.5$ from bottom to top.
 Figure (d) shows a double-log plot of (c), and the dots show $\tau_{\mathrm{c}0}$ and $\tau_{\mathrm{c}}$.}
 \label{Supplement-Fig-rho-off}
\end{figure}

In the presence of nonzero $J$, the Liouvillian mixes the diagonal and off-diagonal elements of the density matrix.
As a result, the Liouvillian eigenmodes that control the slow relaxation of the system have nonzero off-diagonal elements. 
Thus, we expect the slowing down of relaxation processes of the off-diagonal elements.
We vindicate this expectation by numerically solving the Lindblad master equation.
As an initial state, we take the following pure state localized near the left boundary at $l=1$:
\begin{equation}
| \psi_{\mathrm{ini}} \rangle = \sum_{l = 1}^L \psi_{\mathrm{ini},l} | l \rangle, \:\:\:\:\: \psi_{\mathrm{ini},l} = \mathcal{N} \exp \left[ - \frac{(l-1)^2}{4 \sigma^2} \right],
\end{equation}
where $\sigma$ is the standard deviation of the density profile $|\psi_{\mathrm{ini},l}|^2$ and $\mathcal{N}$ is a normalization constant that satisfies $\sum_l |\psi_{\mathrm{ini},l}|^2=1$.
The density matrix is given by $\hat{\rho}_{\mathrm{ini}} = | \psi_{\mathrm{ini}} \rangle \langle \psi_{\mathrm{ini}} |$, whose matrix elements read $\rho_{\mathrm{ini},lm} = \langle l | \hat{\rho}_{\mathrm{ini}} | m \rangle = \psi_{\mathrm{ini},l} \psi_{\mathrm{ini},m}^*$.
We define the amplitude of the off-diagonal elements by
\begin{equation}
\kappa \equiv \sum_{l,m \:(l \neq m)}^L |\rho_{lm}|.
\end{equation}

Figures \ref{Supplement-Fig-rho-off} (a) and (b) show the time evolution of $\kappa$ for different system sizes.
In an early stage, $\kappa$ exhibits a power-law decay up to some crossover timescale $\tau_{\mathrm{c}}$, at which the wave packet of a particle reaches the right boundary and $\kappa$ starts to increase toward a stationary value.
From Fig.~\ref{Supplement-Fig-rho-off} (a), one can see that $\tau_{\mathrm{c}}$ is proportional to the system size.
Since the relaxation time $\tau$ should be larger than $\tau_{\mathrm{c}}$, we find that $\tau$ diverges linearly with increasing the system size.
Figures \ref{Supplement-Fig-rho-off} (c) and (d) show the time evolution of $\kappa$ for different transfer amplitudes $J$.
As mentioned above, for $J=0$, $\kappa$ decays exponentially with rate $\gamma_{\mathrm{R}}+\gamma_{\mathrm{L}}$.
For small but nonzero $J$, one can distinguish three different regimes separated by two crossover timescales $\tau_{\mathrm{c}0}$ and $\tau_{\mathrm{c}}$.
For $t<\tau_{\mathrm{c}0}$, $\kappa$ shows an exponential decay similar to the case of $J=0$.
For an intermediate regime $\tau_{\mathrm{c}0}<t<\tau_{\mathrm{c}}$, $\kappa$ shows a power-law decay $t^{-\alpha}$ with $\alpha \simeq 0.5$.
While $\tau_{\mathrm{c}}$ is independent of $J$, $\tau_{\mathrm{c}0}$ increases with decreasing $J$, and the power-law regime disappears in the $J \to 0$ limit.

\subsection{S3. \ Laser-assisted hopping with spontaneous emission}

In this section, we consider two schemes for physical implementation of a dissipative tight-binding model with an asymmetric stochastic hopping.
Specifically, we propose two types of laser-assisted hopping for ultracold atoms in an optical lattice.
The laser-assisted hopping has been employed to control the phase of a coherent hopping amplitude of atoms, which is known as a synthetic gauge field \cite{Jaksch-03, Aidelsburger-13, Miyake-13}.
This method can also be applied to produce a stochastic hopping between neighboring sites of an optical lattice.
An asymmetric hopping can be realized by using either a single laser field with spatially varying intensity or two laser fields with different frequencies, which selectively induce hopping of a particle to different directions. 
In the following, we discuss microscopic details of these two setups.

\subsubsection{A. \ Asymmetric hopping induced by a spatially varying laser field}

\begin{figure}
 \centering
 \includegraphics[width=0.9\textwidth]{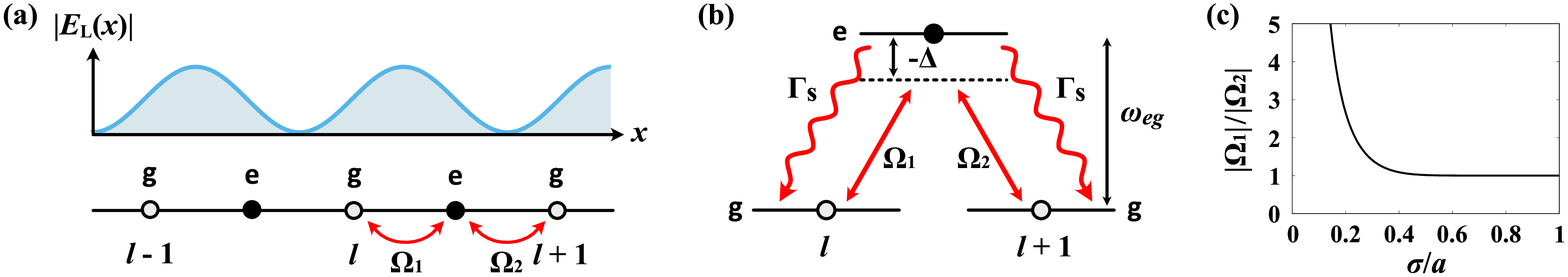}
 \caption{(a) Schematic illustration of the setup for physical implementation.
 The empty and filled circles represent the potential minima for the atom in the ground state $|g\rangle$ and the excited state $|e\rangle$, respectively.
 The upper graph illustrates the intensity of a laser field, which induces the transition between the ground and excited states, as a function of the spatial coordinate.
 (b) Level diagram for atoms at neighboring sites.
 The straight arrows represent the Rabi coupling between $|g\rangle$ and $|e\rangle$ induced by a laser.
 The wavy arrows represent spontaneous emission with rate $\Gamma_{\mathrm{s}}$, and $\Delta=\omega_{\mathrm{L}}-\omega_{eg}$ is the detuning of the laser.
 (c) Ratio between $|\Omega_1|$ and $|\Omega_2|$ given by Eq.~(\ref{Supplement_ratio_Omega_1_Omega_2}).}
 \label{Supplement-laser-assisted-hopping-1}
\end{figure}

Suppose that atoms with two internal states, the ground state $|g \rangle$ and an excited state $|e \rangle$, are trapped in a one-dimensional optical lattice.
Atoms with different internal states experience different lattice potentials, which are shifted from each other by half a period (see Fig.~\ref{Supplement-laser-assisted-hopping-1} (a)).
A stimulated transition between the two internal states is induced by a laser field with spatially varying intensity $|E_{\mathrm{L}}(\mathbf{r})|$ and frequency $\omega_{\mathrm{L}}$.
The total Hamiltonian for the atoms coupled to the photon field is given by
\begin{equation}
\hat{H} = \hat{H}_{\mathrm{A}} + \hat{H}_{\mathrm{F}} + \hat{H}_{\mathrm{A-L}} + \hat{H}_{\mathrm{A-F}},
\end{equation}
where $\hat{H}_{\mathrm{A}}$ and $\hat{H}_{\mathrm{F}}$ denote the Hamiltonians for the atom and the photon field, $\hat{H}_{\mathrm{A-L}}$ is the interaction Hamiltonian between the atom and a coherent laser field, which is treated as a classical field, and $\hat{H}_{\mathrm{A-F}}$ denotes the coupling between the atom and the photon field.

To demonstrate an asymmetric stochastic hopping, it is sufficient to focus on a single atom.
The Hamiltonian for the atom is given by
\begin{equation}
\hat{H}_{\mathrm{A}} = \frac{\hat{p}^2}{2m} + V_g(\hat{x}) |g \rangle \langle g| + V_e(\hat{x}) |e \rangle \langle e| + \frac{\omega_{eg}}{2} (|e \rangle \langle e| - |g \rangle \langle g|),
\label{Supplement_H_A}
\end{equation}
where $V_g(x)=-V_{g0} \cos(2\pi x/a)$ and $V_e(x)=-V_{e0} \cos(2\pi x/a-\pi)$ ($V_{g0}, V_{e0} >0$) are the potentials of an optical lattice for the atom in the ground and excited states, respectively, and $\omega_{eg}$ denotes the excitation energy.
The Planck constant $\hbar$ is set to unity.
Let $\hat{a}_{\mathbf{k} \lambda}^{\dag}$ and $\hat{a}_{\mathbf{k} \lambda}$ be the creation and annihilation operators of a photon with wavevector $\mathbf{k}$ and polarization $\lambda$.
The Hamiltonian of the photon field reads
\begin{equation}
\hat{H}_{\mathrm{F}} = \sum_{\mathbf{k}, \lambda} ck \hat{a}_{\mathbf{k} \lambda}^{\dag} \hat{a}_{\mathbf{k} \lambda}.
\end{equation}

The atomic dipole operator is given by
\begin{equation}
\hat{\mathbf{D}} = \mathbf{M} ( |g \rangle \langle e|+|e \rangle \langle g| ),
\end{equation}
where $\mathbf{M}$ is the electric-dipole matrix element.
The interaction between the atom and a laser field is written as
\begin{equation}
\hat{H}_{\mathrm{A-L}} = - \hat{\mathbf{D}} \cdot \mathbf{E}_{\mathrm{L}}(\hat{\mathbf{r}},t) = - \hat{\mathbf{D}} \cdot (\mathbf{E}_{\mathrm{L}}(\hat{\mathbf{r}})e^{-i\omega_{\mathrm{L}}t} + \mathbf{E}_{\mathrm{L}}^{*}(\hat{\mathbf{r}})e^{i\omega_{\mathrm{L}}t}),
\label{Supplement_H_A_L}
\end{equation}
where $\hat{\mathbf{r}}$ is the position operator of an atom.
With the rotating-wave approximation, Eq.~(\ref{Supplement_H_A_L}) reduces to
\begin{equation}
\hat{H}_{\mathrm{A-L}} \simeq - \mathbf{M} \cdot \mathbf{E}_{\mathrm{L}}(\hat{\mathbf{r}}) |e \rangle \langle g| e^{-i\omega_{\mathrm{L}}t} - \mathbf{M} \cdot \mathbf{E}_{\mathrm{L}}^{*}(\hat{\mathbf{r}}) |g \rangle \langle e| e^{i\omega_{\mathrm{L}}t}.
\label{Supplement_H_A_L_RWA}
\end{equation}
Finally, $\hat{H}_{\mathrm{A-F}}$ is obtained by replacing the classical field $\mathbf{E}_{\mathrm{L}}(\hat{\mathbf{r}},t)$ in Eq.~(\ref{Supplement_H_A_L}) with the quantized field as
\begin{equation}
\hat{H}_{\mathrm{A-F}} = \sum_{\mathbf{k}, \lambda} g_{\mathbf{k} \lambda} (|e \rangle \langle g| \hat{a}_{\mathbf{k} \lambda} e^{i \mathbf{k} \cdot \hat{\mathbf{r}}} + |g \rangle \langle e| \hat{a}_{\mathbf{k}\lambda}^{\dag} e^{-i \mathbf{k} \cdot \hat{\mathbf{r}}} ),
\end{equation}
where $g_{\mathbf{k} \lambda}$ is the coupling strength between an atom and the photon field with wavevector $\mathbf{k}$ and polarization $\lambda$.

By eliminating the degrees of freedom of photons, one can obtain the quantum master equation for the atom \cite{Dalibard-85, Pichler-10}.
If one focuses on the internal degrees of freedom of the atom, the coupling to photons leads to an optical Bloch equation:
\begin{equation}
\biggl( \frac{d \hat{\rho}}{dt} \biggr)_{\mathrm{s}}= \Gamma_{\mathrm{s}} \biggl( |g \rangle \langle e| \hat{\rho} |e \rangle \langle g| - \frac{1}{2} |e \rangle \langle e| \hat{\rho} - \frac{1}{2} \hat{\rho} |e \rangle \langle e| \biggr),
\label{Supplement_Bloch_eq}
\end{equation}
where $(...)_{\mathrm{s}}$ represents the contribution to the time evolution resulting from spontaneous emission of photons, whose rate is given by $\Gamma_{\mathrm{s}}$.
We have ignored the recoil of the atom in spontaneous emission process.

We assume that the optical lattice $V_{g(e)}(x)$ is so deep that tunneling between neighboring sites is negligible.
Thus, for the atom in $|g \rangle$ to move to its neighboring site, it must be excited to $|e \rangle$ by a laser field $\mathbf{E}_{\mathrm{L}}(\mathbf{r})$ (see Fig.~\ref{Supplement-laser-assisted-hopping-1} (b)).
As shown in Fig.~\ref{Supplement-laser-assisted-hopping-1} (a), the laser intensity $|\mathbf{E}_{\mathrm{L}}(\mathbf{r})|$ is assumed to vary periodically in the $x$ direction with the same period as that of the optical lattice.
From Eq.~(\ref{Supplement_H_A_L_RWA}), the Rabi frequencies corresponding to transitions from the ground states at site $l$ or site $l+1$ to the excited state at site $l$ are given by
\begin{equation}
|\Omega_1| = c \int d\mathbf{r} |\mathbf{E}_{\mathrm{L}}(\mathbf{r})| w_e(\mathbf{r}-\mathbf{R}_l^{(e)}) w_g(\mathbf{r}-\mathbf{R}_l^{(g)}),
\label{Supplement_Omega_1}
\end{equation}
\begin{equation}
|\Omega_2| = c \int d\mathbf{r} |\mathbf{E}_{\mathrm{L}}(\mathbf{r})| w_e(\mathbf{r}-\mathbf{R}_l^{(e)}) w_g(\mathbf{r}-\mathbf{R}_{l+1}^{(g)}),
\label{Supplement_Omega_2}
\end{equation}
where $\mathbf{R}_l^{(g)}$ and $\mathbf{R}_l^{(e)}$ denote the positions of the ground and excited states at site $l$, respectively, $w_{g(e)}(\mathbf{r})$ is the Wannier function for the ground (excited) state, and $c$ is a constant independent of the laser intensity $|\mathbf{E}_{\mathrm{L}}(\mathbf{r})|$.

Let $\hat{b}_l$ and $\hat{c}_l$ be the annihilation operators of the ground and excited states at site $l$, respectively.
In the frame rotating at frequency $\omega_{\mathrm{L}}$, the tight-binding Hamiltonian with the internal degrees of freedom is given by
\begin{equation}
\hat{H}_{eg} = - \Omega_1 \sum_l (\hat{c}_l^{\dag} \hat{b}_l + \hat{b}_l^{\dag} \hat{c}_l) - \Omega_2 \sum_l (\hat{b}_{l+1}^{\dag} \hat{c}_l + \hat{c}_l^{\dag} \hat{b}_{l+1}) - \Delta \sum_l \hat{c}_l^{\dag} \hat{c}_l,
\label{Supplement_H_eg}
\end{equation}
where $\Delta=\omega_{\mathrm{L}}-\omega_{eg}$ is the detuning of the laser.
From Eq.~(\ref{Supplement_Bloch_eq}), the quantum master equation reads
\begin{equation}
\frac{d \hat{\rho}}{dt} = -i[\hat{H}_{eg}, \hat{\rho}] + \sum_{\nu = \mathrm{R},\mathrm{L}} \sum_l \biggl( \hat{M}_{\nu,l} \hat{\rho} \hat{M}_{\nu,l}^{\dag} - \frac{1}{2} \bigl\{ \hat{\rho}, \hat{M}_{\nu,l}^{\dag} \hat{M}_{\nu,l} \bigr\} \biggr),
\end{equation}
where the Lindblad operators are given by
\begin{equation}
\hat{M}_{\mathrm{R},l} = \sqrt{\Gamma_{\mathrm{s}}} \hat{b}_{l+1}^{\dag} \hat{c}_l, \:\:\: \hat{M}_{\mathrm{L},l} = \sqrt{\Gamma_{\mathrm{s}}} \hat{b}_{l}^{\dag} \hat{c}_l.
\end{equation}

When the detuning is sufficiently large ($|\Delta| \gg |\Omega|, \Gamma_{\mathrm{s}}$), the excited state can be eliminated adiabatically to obtain a quantum master equation for the degrees of freedom of the ground state only.
The resulting master equation reads
\begin{equation}
\frac{d \hat{\rho}}{dt} = -i[\hat{H}_{\mathrm{TB}}, \hat{\rho}] + \sum_{\alpha = \mathrm{H},\mathrm{D}} \sum_{\nu = \mathrm{R},\mathrm{L}} \sum_l \biggl( \hat{L}_{\nu,l}^{(\alpha)} \hat{\rho} \hat{L}_{\nu,l}^{(\alpha) \dag} - \frac{1}{2} \bigl\{ \hat{\rho}, \hat{L}_{\nu,l}^{(\alpha) \dag} \hat{L}_{\nu,l}^{(\alpha)} \bigr\} \biggr),
\label{Supplement_master_eq_TB}
\end{equation}
where $\hat{H}_{\mathrm{TB}}$ is the tight-binding Hamiltonian
\begin{equation}
\hat{H}_{\mathrm{TB}} = - J \sum_l (\hat{b}_{l+1}^{\dag} \hat{b}_l + \hat{b}_l^{\dag} \hat{b}_{l+1}),
\end{equation}
with a tunneling amplitude
\begin{equation}
J = \frac{\Omega_1 \Omega_2}{|\Delta|}.
\label{Supplement_J_eff}
\end{equation}
The Lindblad operators $\hat{L}_{\nu,l}^{(\alpha)}$ are given by
\begin{equation}
\hat{L}_{\mathrm{R},l}^{(\mathrm{H})} = \frac{\sqrt{\Gamma_{\mathrm{s}}} \Omega_{1}}{\Delta} \hat{b}_{l+1}^{\dag} \hat{b}_l, \:\:\: \hat{L}_{\mathrm{L},l}^{(\mathrm{H})} = \frac{\sqrt{\Gamma_{\mathrm{s}}} \Omega_{2}}{\Delta} \hat{b}_{l-1}^{\dag} \hat{b}_l,
\label{Supplement_L_hopping}
\end{equation}
\begin{equation}
\hat{L}_{\mathrm{R},l}^{(\mathrm{D})} = \frac{\sqrt{\Gamma_{\mathrm{s}}} \Omega_{1}}{\Delta} \hat{b}_l^{\dag} \hat{b}_l, \:\:\: \hat{L}_{\mathrm{L},l}^{(\mathrm{D})} = \frac{\sqrt{\Gamma_{\mathrm{s}}} \Omega_{2}}{\Delta} \hat{b}_l^{\dag} \hat{b}_l.
\label{Supplement_L_dephasing}
\end{equation}
Here, $\hat{L}_{\mathrm{R}(\mathrm{L}),l}^{(\mathrm{H})}$ represents a stochastic hopping to the right (left) neighboring site via the excited state, and $\hat{L}_{\mathrm{R}(\mathrm{L}),l}^{(\mathrm{D})}$ represents an on-site dephasing process resulting from a backward process from the right (left) excited state to the original ground state.
The rates of the stochastic hopping are given by
\begin{equation}
\gamma_{\mathrm{R}} = \frac{\Gamma_{\mathrm{s}} \Omega_{1}^2}{\Delta^2}, \:\:\: \gamma_{\mathrm{L}} = \frac{\Gamma_{\mathrm{s}} \Omega_{2}^2}{\Delta^2}.
\label{Supplement_gamma_R_L}
\end{equation}
The strength of the on-site dephasing reads
\begin{equation}
\gamma_{\mathrm{d}} = \gamma_{\mathrm{R}} + \gamma_{\mathrm{L}} = \frac{\Gamma_{\mathrm{s}} \Omega_{1}^2}{\Delta^2} + \frac{\Gamma_{\mathrm{s}} \Omega_{2}^2}{\Delta^2}.
\end{equation}
From Eqs.~(\ref{Supplement_J_eff}) and (\ref{Supplement_gamma_R_L}), a typical value of the stochastic hopping rate $\gamma_{\mathrm{R}(\mathrm{L})}$ is $\Gamma_{\mathrm{s}}J/|\Delta|$.

To understand how to control the asymmetry of hopping, let us calculate $|\Omega_1|$ and $|\Omega_2|$ explicitly as functions of the spatial phase of $|\mathbf{E}_{\mathrm{L}}(\mathbf{r})|$.
For simplicity, we consider a one-dimensional system with $V_{g0}=V_{e0}$, which implies $w_g(x)=w_e(x)=w(x)$.
Then, Eqs.~(\ref{Supplement_Omega_1}) and (\ref{Supplement_Omega_2}) reduce to
\begin{equation}
|\Omega_1| = c \int dx |\mathbf{E}_{\mathrm{L}}(x)| w(x-a/2) w(x),
\label{Supplement_Omega_1_1D}
\end{equation}
\begin{equation}
|\Omega_2| = c \int dx |\mathbf{E}_{\mathrm{L}}(x)| w(x-a/2) w(x-a),
\label{Supplement_Omega_2_1D}
\end{equation}
where the position of the ground state at site $l$ is set to the origin by using the periodicity of $|\mathbf{E}_{\mathrm{L}}(x)|$.
We assume that the intensity of the laser field is given by
\begin{equation}
|\mathbf{E}_{\mathrm{L}}(x)| = E_0 \left[ 1+\cos \left( \frac{2\pi x}{a} - \phi \right) \right],
\label{Supplement_E_L}
\end{equation}
where $\phi$ is its spatial phase.
If the optical lattice is so deep that the particle motion around its minima is described by a harmonic oscillator, the Wannier function is given by a Gaussian function,
\begin{equation}
w(x) = (2\pi \sigma^2)^{-1/4} \exp \left( - \frac{x^2}{4\sigma^2} \right),
\label{Supplement_Wannier}
\end{equation}
where the width $\sigma$ of the wavefunction can be controlled by changing the depth of the optical lattice.
By substituting Eqs.~(\ref{Supplement_E_L}) and (\ref{Supplement_Wannier}) into Eqs.~(\ref{Supplement_Omega_1_1D}) and (\ref{Supplement_Omega_2_1D}), $|\Omega_1|$ and $|\Omega_2|$ can be calculated as
\begin{equation}
|\Omega_1| = c E_0 \exp \left(-\frac{a^2}{32\sigma^2}\right) \left[ 1+ \exp \left(-\frac{2\pi^2 \sigma^2}{a^2}\right) \cos \left( \phi-\frac{\pi}{2} \right) \right],
\label{Supplement_Omega_1_expression}
\end{equation}
\begin{equation}
|\Omega_2| = c E_0 \exp \left(-\frac{a^2}{32\sigma^2}\right) \left[ 1+ \exp \left(-\frac{2\pi^2 \sigma^2}{a^2}\right) \cos \left( \phi+\frac{\pi}{2} \right) \right].
\label{Supplement_Omega_2_expression}
\end{equation}
From Eqs.~(\ref{Supplement_Omega_1_expression}) and (\ref{Supplement_Omega_2_expression}), we find $|\Omega_1|=|\Omega_2|$ for $\phi=0$ and $\pi$.
The ratio between $|\Omega_1|$ and $|\Omega_2|$ attains its maximum for $\phi=\pi/2$ as in Fig.~\ref{Supplement-laser-assisted-hopping-1} (a), where the ratio between them reads
\begin{equation}
\frac{|\Omega_1|}{|\Omega_2|} = \frac{1+\exp(-2\pi^2 \sigma^2/a^2)}{1-\exp(-2\pi^2 \sigma^2/a^2)}.
\label{Supplement_ratio_Omega_1_Omega_2}
\end{equation}
Figure \ref{Supplement-laser-assisted-hopping-1} (c) shows $|\Omega_1|/|\Omega_2|$ for $\phi=\pi/2$ as a function of $\sigma/a$.
For $\sigma/a \ll 1$, we have $|\Omega_1|/|\Omega_2| \propto a^2/\sigma^2$, and $|\Omega_1|=|\Omega_2|$ in the limit $\sigma/a \to \infty$.
In other words, if the depth of the optical lattice increases ($\sigma \to 0$), the ratio $|\Omega_1|/|\Omega_2|$ diverges while both $|\Omega_1|$ and $|\Omega_2|$ exponentially vanish.

\subsubsection{B. \ Asymmetric hopping induced by two laser fields with different frequencies}

\begin{figure}
 \centering
 \includegraphics[width=0.3\textwidth]{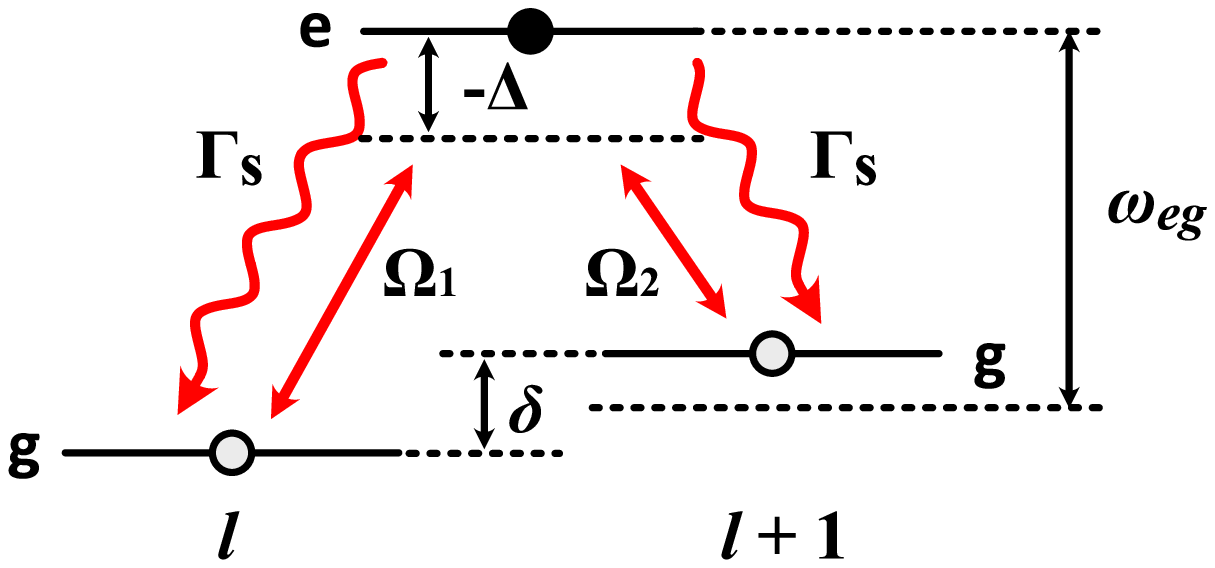}
 \caption{Level diagram for atoms at neighboring sites.
 The straight arrows represent the Rabi coupling between $|g\rangle$ and $|e\rangle$ induced by two lasers with a frequency difference $\omega_{\mathrm{L}1}-\omega_{\mathrm{L}2}=\delta$.
 The wavy arrows represent spontaneous emission with rate $\Gamma_{\mathrm{s}}$, and $\delta$ is an energy offset between neighboring sites.}
 \label{Supplement-laser-assisted-hopping-2}
\end{figure}

Asymmetric Rabi couplings ($|\Omega_1|\neq|\Omega_2|$) can also be realized by using two laser fields with different frequencies, which drive hopping to different directions \cite{Jaksch-03, Aidelsburger-13, Miyake-13}.
In addition to the Hamiltonian given by Eq.~(\ref{Supplement_H_A}), we introduce a potential energy with a constant gradient, $\hat{H}_{\mathrm{grad}}=(\delta/a)x$, where $\delta$ is an energy offset between adjacent sites.
Two laser fields with spatially uniform intensities are given by
\begin{eqnarray}
\mathbf{E}_{\mathrm{L}}^{(1)}(\mathbf{r},t) = \mathbf{E}_0^{(1)} e^{-i\omega_{\mathrm{L}1}t} + \mathbf{E}_0^{(1)*} e^{i\omega_{\mathrm{L}1}t}, \:\:\:\:\:
\mathbf{E}_{\mathrm{L}}^{(2)}(\mathbf{r},t) = \mathbf{E}_0^{(2)} e^{-i\omega_{\mathrm{L}2}t} + \mathbf{E}_0^{(2)*} e^{i\omega_{\mathrm{L}2}t},
\end{eqnarray}
where $\omega_{\mathrm{L}1}=\omega_{eg}+\delta/2+\Delta$ and $\omega_{\mathrm{L}2}=\omega_{eg}-\delta/2+\Delta$.
The laser fields $1$ and $2$ drive transitions from the ground states at sites $l$ and $l+1$ to the excited state, respectively (see Fig.~\ref{Supplement-laser-assisted-hopping-2}).
The Rabi couplings $|\Omega_1|$ and $|\Omega_2|$ are given by
\begin{equation}
|\Omega_1| = c |\mathbf{E}_{0}^{(1)}| \int d\mathbf{r} w_e(\mathbf{r}-\mathbf{R}_l^{(e)}) w_g(\mathbf{r}-\mathbf{R}_l^{(g)}),
\end{equation}
\begin{equation}
|\Omega_2| = c |\mathbf{E}_{0}^{(2)}| \int d\mathbf{r} w_e(\mathbf{r}-\mathbf{R}_l^{(e)}) w_g(\mathbf{r}-\mathbf{R}_{l+1}^{(g)}).
\end{equation}
It should be noted that while the laser $2$ ($1$) also contributes to the transition from the ground state at site $l$ ($l+1$) to the excited state, such a contribution is negligible when $\Omega_1, \Omega_2 \ll \delta$.
In the rotating frame with the frequency $\omega_{eg}$, we obtain a tight-binding Hamiltonian given by Eq.~(\ref{Supplement_H_eg}).
The potential term $\hat{H}_{\mathrm{grad}}$ does not appear in the tight-binding Hamiltonian because it is canceled by a term resulting from switching to the rotating frame.
After elimination of the excited state, the master equation and the Lindblad operators are given by Eqs.~(\ref{Supplement_master_eq_TB}), (\ref{Supplement_L_hopping}), and (\ref{Supplement_L_dephasing}).

\subsection{S4. \ Effect of the on-site dephasing on the Liouvillian skin effect}

As shown in Sec.~S3, the laser-assisted hopping with spontaneous emission gives rise to an additional on-site dephasing.
In this section, we demonstrate that such an on-site dephasing does not change the qualitative features of the prototypical model discussed in the main text.
We consider the Lindblad master equation 
\begin{equation}
\frac{d \hat{\rho}}{dt} = -i[\hat{H},\hat{\rho}] + \sum_{\alpha=\mathrm{R},\mathrm{L},\mathrm{d}} \sum_{l} \left( \hat{L}_{\alpha,l} \hat{\rho} \hat{L}_{\alpha,l}^{\dag} - \frac{1}{2} \{ \hat{L}_{\alpha,l}^{\dag} \hat{L}_{\alpha,l}, \hat{\rho} \} \right) \equiv \mathcal{L}(\hat{\rho}),
\label{Supplement_Master_eq}
\end{equation}
where the Lindblad operators are given by $\hat{L}_{\mathrm{R},l}=\sqrt{\gamma_{\mathrm{R}}}\hat{b}_{l+1}^{\dag}\hat{b}_{l}$, $\hat{L}_{\mathrm{L},l}=\sqrt{\gamma_{\mathrm{L}}}\hat{b}_{l-1}^{\dag}\hat{b}_{l}$, and $\hat{L}_{\mathrm{d},l}=\sqrt{\gamma_{\mathrm{d}}}\hat{b}_{l}^{\dag}\hat{b}_{l}$.
In the setup discussed in Sec.~S3, the strength of the on-site dephasing and the rates of stochastic hopping are related to each other by $\gamma_{\mathrm{d}}=\gamma_{\mathrm{R}}+\gamma_{\mathrm{L}}$.
However, we treat them as free parameters in the following calculations.

First, it should be noted that, for the case of $J=0$, the on-site dephasing does not affect the eigenmodes.
In fact, one can confirm that the action of $\mathcal{L}$ to the diagonal basis $|l \rangle \langle l|$ is given by the same equation with $\gamma_{\mathrm{d}}=0$, and the action to the off-diagonal basis is given by $\mathcal{L}(|l \rangle \langle m|)=-(\gamma_{\mathrm{R}}+\gamma_{\mathrm{L}}+\gamma_{\mathrm{d}})|l \rangle \langle m|, \:(l \neq m)$.
In other words, the on-site dephasing does not change the eigenvalues for the diagonal eigenmodes but shifts those for the off-diagonal eigenmodes by $-\gamma_{\mathrm{d}}$.
In the presence of nonzero $J$, since the diagonal and off-diagonal subspaces are coupled to each other by coherent hopping, the on-site dephasing has a nontrivial effect on the eigenmodes and eigenvalues.

\begin{figure}
 \centering
 \includegraphics[width=0.6\textwidth]{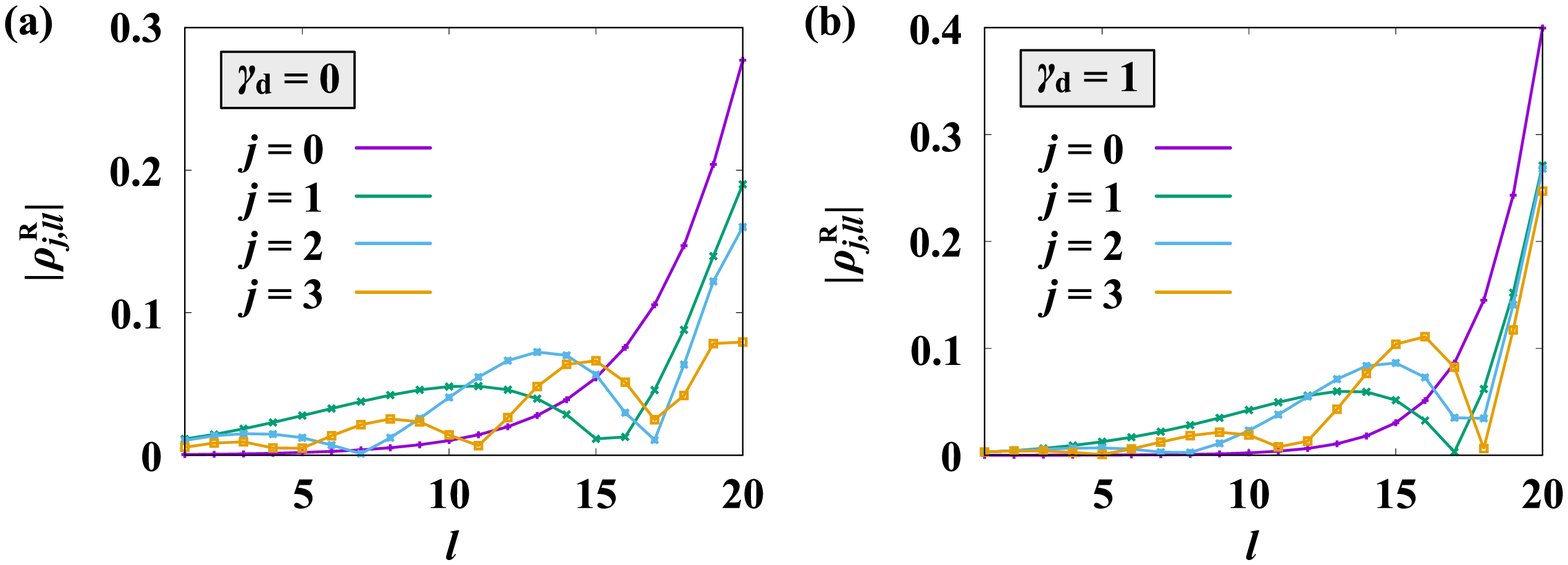}
 \caption{Diagonal elements of the right eigenmodes $|\rho_{j,ll}^{\mathrm{R}}|$ with $J=\gamma_{\mathrm{R}}=1$ and $\gamma_{\mathrm{L}}=0.2$, and the system size $L=20$.
 The strength of the on-site dephasing is $\gamma_{\mathrm{d}}=0$ for (a) and $\gamma_{\mathrm{d}}=1$ for (b).
 All eigenmodes are normalized as $\| \hat{\rho}_{j}^{\mathrm{R}} \|_{\mathrm{tr}}=1$.}
 \label{Supplement-Fig-eigvec-deph}
\end{figure}

\begin{figure}
 \centering
 \includegraphics[width=0.45\textwidth]{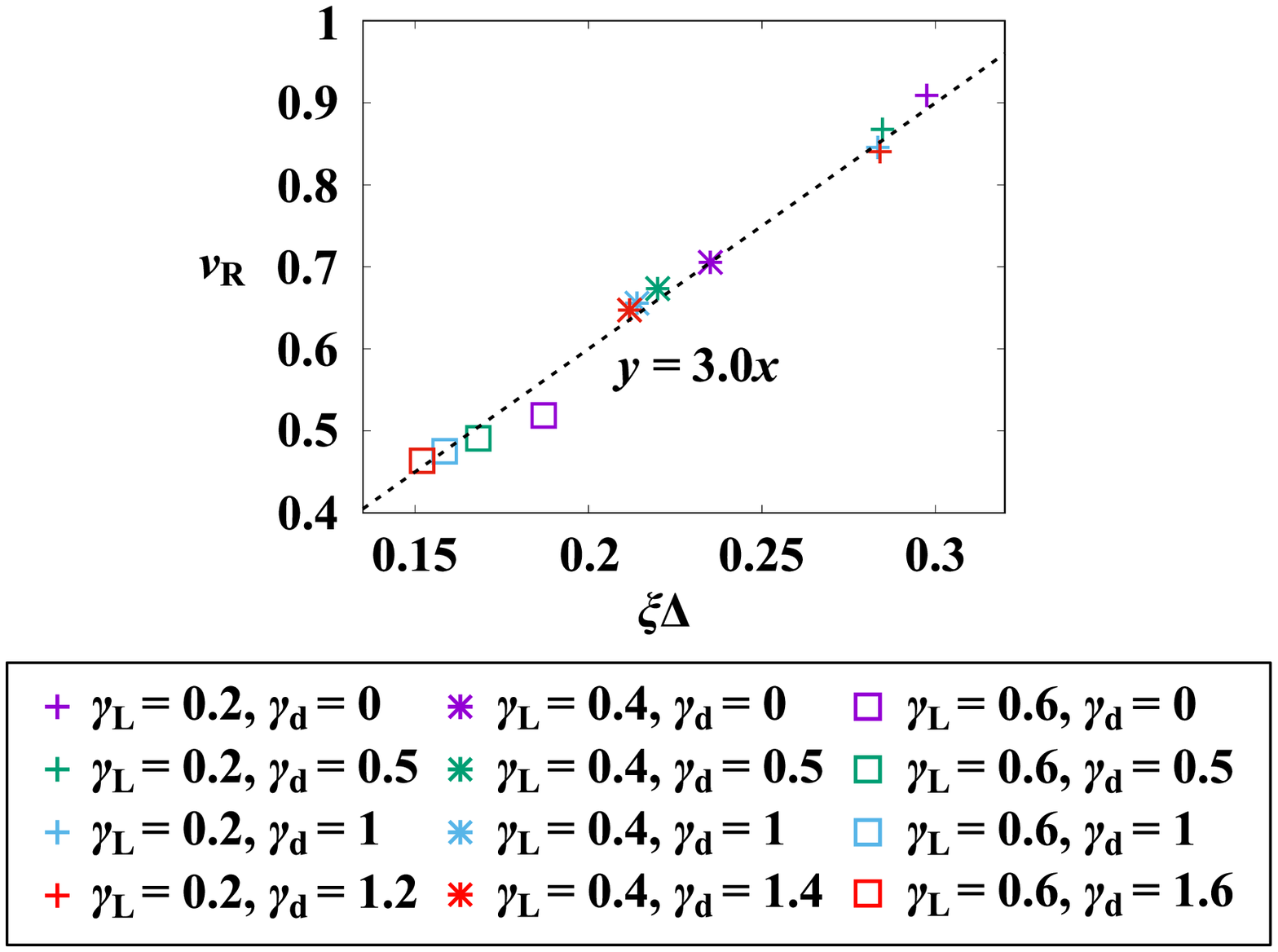}
 \caption{Plots of $v_{\mathrm{R}}$ versus $\xi\Delta$ with $J=\gamma_{\mathrm{R}}=1$, and $\gamma_{\mathrm{L}}=0.2$ (cross), $0.4$ (star), $0.6$ (square), and $\gamma_{\mathrm{d}}=0$ (purple), $0.5$ (green), $1$ (blue), $\gamma_{\mathrm{R}}+\gamma_{\mathrm{L}}$ (red) for the system size $L=40$.
 The dashed line represents the least squares fitting by $y=ax$.}
 \label{Supplement-Fig-gap-xi-vR}
\end{figure}

Figure \ref{Supplement-Fig-eigvec-deph} shows the diagonal elements of the right eigenmodes $|\rho_{j,ll}^{\mathrm{R}}|$ for (a) $\gamma_{\mathrm{d}}=0$ and (b) $\gamma_{\mathrm{d}}=1$.
One can confirm that the localization length $\xi$ decreases as $\gamma_{\mathrm{d}}$ increases; in other words, the on-site dephasing enhances the skin effect.
This is because the on-site dephasing suppresses the coherent hopping of the particle.
Figure \ref{Supplement-Fig-gap-xi-vR} shows $v_{\mathrm{R}}=L/\tau$ and $\xi\Delta$ for several sets of parameters, which include those satisfying $\gamma_{\mathrm{d}}=\gamma_{\mathrm{R}}+\gamma_{\mathrm{L}}$.
The relaxation time $\tau$ is determined from the time evolution of the density $n_L(t)$ at the right boundary starting from the initial state completely localized at the left boundary.
From Fig.~\ref{Supplement-Fig-gap-xi-vR}, we can conclude that Eq.~(9) also holds in the presence of on-site dephasing.

\section{S5. \ Absence of Liouvillian skin effect under the detailed balance condition}

The Liouvillian skin effect is a nonequilibrium effect that is unique to dissipative systems driven far from equilibrium.
In this section, we show that in dissipative systems weakly coupled to a thermal environment, the right and left eigenmodes of the Liouvillian cannot be localized near opposite boundaries of the system.
This means that the overlap between the right and left eigenmodes cannot be exponentially small.

We consider a system weakly coupled to a thermal environment with inverse temperature $\beta$.
In the following argument, the Hamiltonian $\hat{H}$ is assumed to be spatially uniform, but dissipation is not necessarily uniform.
Thus, this proof is also applicable to boundary-dissipative systems.
The steady state $\hat{\sigma}$ is given by the Gibbs distribution:
$\hat{\sigma} = e^{-\beta \hat{H}}/\mathrm{Tr}[e^{-\beta \hat{H}}]$.
We define a superoperator $\Gamma_{\sigma}$ by $\Gamma_{\sigma}(\hat{\rho})=\hat{\sigma} \hat{\rho}$ for a given operator $\hat{\rho}$.
The time reversal operation is described by an anti-unitary operator $\hat{\theta}$, which satisfies $\hat{\theta}^2=1$ and $\hat{\theta}=\hat{\theta}^{-1}=\hat{\theta}^{\dag}$.
The time reversal superoperator $\mathcal{T}$ acting on an operator $\hat{\rho}$ is defined by $\mathcal{T}(\hat{\rho})=\hat{\theta} \hat{\rho} \hat{\theta}^{-1}$.
$\mathcal{T}$ is also an anti-unitary superoperator, and satisfies $\mathcal{T}^2=1$ and $\mathcal{T}=\mathcal{T}^{-1}=\mathcal{T}^{\dag}$.
The condition of micro-reversibility or detailed balance for the Liouvillian $\mathcal{L}$ is given by the following relation \cite{Chetrite-13}:
\begin{equation}
\mathcal{T} \Gamma_{\sigma} \mathcal{L}^{\dag} = \mathcal{L} \mathcal{T} \Gamma_{\sigma}.
\label{Supplement_detailed_balance_1}
\end{equation}
For the case of a spinless particle, $\mathcal{T}$ is equivalent to the complex conjugation operator, $\mathcal{T}(\hat{\rho})=\hat{\rho}^*$, and therefore Eq.~\eqref{Supplement_detailed_balance_1} reads
\begin{equation}
\hat{\sigma} \mathcal{L}^{\dag}(\hat{\rho})^* = \mathcal{L}(\hat{\sigma} \hat{\rho}^*)
\label{Supplement_detailed_balance_2}
\end{equation}
for any operator $\hat{\rho}$.
It can be verified that the so-called Davies generator, which describes a quantum dissipative system weakly coupled to a thermal environment, satisfies the detailed balance condition \cite{Davies-79, Spohn-78}.
Note that Eq.~\eqref{Supplement_detailed_balance_1} and its conjugate imply $\Gamma_{\sigma} \mathcal{T} = \mathcal{T} \Gamma_{\sigma}$, which leads to $\hat{\theta} \hat{H} \hat{\theta}^{-1}=\hat{H}$.
If one defines $\tilde{\mathcal{L}}:=\Gamma_{\sigma^{1/2}}^{-1} \mathcal{L} \Gamma_{\sigma^{1/2}}$, Eq.~\eqref{Supplement_detailed_balance_1} is rewritten as 
\begin{equation}
\mathcal{T}\tilde{\mathcal{L}}^{\dag}\mathcal{T}^{-1}=\tilde{\mathcal{L}}.
\label{Supplement_detailed_balance_3}
\end{equation}
The class of operators that satisfy Eq.~\eqref{Supplement_detailed_balance_3} is known as $\mathrm{AI}^{\dag}$ in the context of non-Hermitian physics \cite{Kawabata-19}.

The right and left eigenmodes of $\mathcal{L}$ are defined by
\begin{equation}
\mathcal{L}(\hat{\rho}^{\mathrm{R}}_j) = \lambda_j \hat{\rho}^{\mathrm{R}}_j, \quad \mathcal{L}^{\dag}(\hat{\rho}^{\mathrm{L}}_j) = \lambda_j^* \hat{\rho}^{\mathrm{L}}_j.
\end{equation}
By using Eq.~\eqref{Supplement_detailed_balance_1}, we have
\begin{equation}
\mathcal{L}^{\dag} \Gamma_{\sigma}^{-1} \mathcal{T} (\hat{\rho}^{\mathrm{R}}_j) = \Gamma_{\sigma}^{-1} \mathcal{T} \mathcal{L} (\hat{\rho}^{\mathrm{R}}_j) = \lambda_j^* \Gamma_{\sigma}^{-1} \mathcal{T} (\hat{\rho}^{\mathrm{R}}_j),
\end{equation}
which implies that
\begin{equation}
\hat{\rho}^{\mathrm{L}}_j \propto \Gamma_{\sigma}^{-1} \mathcal{T} (\hat{\rho}^{\mathrm{R}}_j),
\end{equation}
or equivalently,
\begin{equation}
\hat{\rho}^{\mathrm{L}}_j \propto \hat{\sigma}^{-1 } (\hat{\rho}^{\mathrm{R}}_j)^*,
\label{Supplement_left_rignt_eigenmode}
\end{equation}
provided that there is no degeneracy.

For simplicity, we consider a single-particle system.
Let $\ket{x}$ be the state in which the particle is located at position $x$.
Then, Eq.~\eqref{Supplement_left_rignt_eigenmode} can be rewritten as
\begin{equation}
\bra{x_1} \hat{\rho}^{\mathrm{R}}_j \ket{x_2}^* \propto \int dy \bra{x_1} \hat{\sigma} \ket{y} \bra{y} \hat{\rho}^{\mathrm{L}}_j \ket{x_2}.
\label{Supplement_left_rignt_eigenmode_element}
\end{equation}
We here assume that the steady state $\hat{\sigma}$ is short-range correlated, which means that $|\bra{x} \hat{\sigma} \ket{y}|$ rapidly decreases to zero as $|x-y|$ increases.
Since the Hamiltonian is uniform, $\bra{x} \hat{\sigma} \ket{x}$ is extended over the entire system.
Then, we have the following results from Eq.~\eqref{Supplement_left_rignt_eigenmode_element}: (i) If $\hat{\rho}^{\mathrm{L}}_j$ is extended over the entire system, so is $\hat{\rho}^{\mathrm{R}}_j$, and (ii) If $\hat{\rho}^{\mathrm{L}}_j$ is localized near a boundary of the system, $\hat{\rho}^{\mathrm{R}}_j$ is also localized near the {\it same} boundary.
Thus, $\hat{\rho}^{\mathrm{R}}_j$ and $\hat{\rho}^{\mathrm{L}}_j$ cannot be localized near opposite boundaries of the system, and therefore the overlap between them cannot be exponentially small by the skin effect.

Note that our proof does not exclude the possibility that an exponentially small overlap between the right and left eigenmodes occurs without the skin effect.
In Ref.~\cite{Mori-20}, the authors investigate a hard-core Bose-Hubbard model under boundary dephasing, which satisfies the detailed balance condition \eqref{Supplement_detailed_balance_1} with an infinite-temperature steady state, and find a superexponentially small overlap $e^{-O(L^2)}$ between the right and left eigenmodes of the Liouvillian.
This behavior cannot be explained by spatial localization of eigenmodes.


\begin{thebibliography}{99}

\bibitem{Bloch-08} I. Bloch, Quantum coherence and entanglement with ultracold atoms in optical lattices, Nature {\bf 453}, 1016 (2008).
\bibitem{Diehl-08} S. Diehl, A. Micheli, A. Kantian, B. Kraus, H. P. B\"{u}chler, and P. Zoller, Quantum states and phases in driven open quantum systems with cold atoms, Nature Phys. {\bf 4}, 878 (2008).
\bibitem{Weimer-08} H. Weimer, R. L\"{o}w, T. Pfau, and H. P. B\"{u}chler, Quantum Critical Behavior in Strongly Interacting Rydberg Gases, Phys. Rev. Lett. {\bf 101}, 250601 (2008).
\bibitem{Tomadin-11} A. Tomadin, S. Diehl, and P. Zoller, Nonequilibrium phase diagram of a driven and dissipative many-body system, Phys. Rev. A {\bf 83}, 013611 (2011).
\bibitem{Lee-11} T. E. Lee, H. H\"{a}ffner, and M. C. Cross, Antiferromagnetic phase transition in a nonequilibrium lattice of Rydberg atoms, Phys. Rev. A {\bf 84}, 031402(R) (2011).
\bibitem{Ludwig-13} M. Ludwig and F. Marquardt, Quantum many-body dynamics in optomechanical arrays,  Phys. Rev. Lett. {\bf 111}, 073603 (2013).

\bibitem{Pellizzari-95} T. Pellizzari, S. A. Gardiner, J. I. Cirac, and P. Zoller, Decoherence, Continuous Observation, and Quantum Computing: A Cavity QED Model, Phys. Rev. Lett. {\bf 75}, 3788 (1995).
\bibitem{Jaksch-00} D. Jaksch, J. I. Cirac, P. Zoller, S. L. Rolston, R. C\^{o}t\'{e}, and M. D. Lukin, Fast Quantum Gates for Neutral Atoms, Phys. Rev. Lett. {\bf 85}, 2208 (2000).
\bibitem{Balasubramanian-09} G. Balasubramanian, P. Neumann, D. Twitchen, M. Markham, R. Kolesov, N. Mizuochi, J. Isoya, J. Achard, J. Beck, J. Tissler, V. Jacques, P. R. Hemmer, F. Jelezko, and J. Wrachtrup, Ultralong spin coherence time in isotopically engineered diamond, Nature Mater. {\bf 8}, 383 (2009).
\bibitem{Isenhower-10} L. Isenhower, E. Urban, X. L. Zhang, A. T. Gill, T. Henage, T. A. Johnson, T. G. Walker, and M. Saffman, Demonstration of a Neutral Atom Controlled-NOT Quantum Gate, Phys. Rev. Lett. {\bf 104}, 010503 (2010).
\bibitem{Weimer-10} H. Weimer, M. M\"{u}ller, I. Lesanovsky, P. Zoller, and H. P. B\"{u}chler,  A Rydberg quantum simulator, Nature Phys. {\bf 6}, 382 (2010).
\bibitem{Lanyon-11} B. P. Lanyon, C. Hempel, D. Nigg, M. M\"{u}ller, R. Gerritsma, F. Z\"{a}hringer, P. Schindler, J. T. Barreiro, M. Rambach, G. Kirchmair, M. Hennrich, P. Zoller, R. Blatt, C. F. Roos, Universal Digital Quantum Simulation with Trapped Ions, Science {\bf 334}, 57 (2011).

\bibitem{Sachdev} S. Sachdev, {\it Quantum Phase Transitions} (Cambridge University Press, Cambridge, 2001).

\bibitem{Cai-13} Z. Cai and T. Barthel, Algebraic versus Exponential Decoherence in Dissipative Many-Particle Systems, Phys. Rev. Lett. {\bf 111}, 150403 (2013).
\bibitem{Bonnes-14} L. Bonnes, D. Charrier, and A. M. L\"{a}uchli, Dynamical and steady-state properties of a Bose-Hubbard chain with bond dissipation: A study based on matrix product operators, Phys. Rev. A {\bf 90}, 033612 (2014).
\bibitem{Znidaric-15} M. \v{Z}nidari\v{c}, Relaxation times of dissipative many-body quantum systems, Phys. Rev. E {\bf 92}, 042143 (2015).

\bibitem{Kessler-12} E. M. Kessler, G. Giedke, A. Imamoglu, S. F. Yelin, M. D. Lukin, and J. I. Cirac, Dissipative phase transition in a central spin system, Phys. Rev. A {\bf 86}, 012116 (2012).
\bibitem{Honing-12} M. H\"{o}ning, M. Moos, and M. Fleischhauer, Critical exponents of steady-state phase transitions in fermionic lattice models, Phys. Rev. A {\bf 86}, 013606 (2012).
\bibitem{Horstmann-13} B. Horstmann, J. I. Cirac, and G. Giedke, Noise-driven dynamics and phase transitions in fermionic systems, Phys. Rev. A {\bf 87}, 012108 (2013).
\bibitem{Minganti-18} F. Minganti, A. Biella, N. Bartolo, and C. Ciuti, Spectral theory of Liouvillians for dissipative phase transitions, Phys. Rev. A {\bf 98}, 042118 (2018).

\bibitem{Levin} D. A. Levin, Y. Peres, and E. L. Wilmer, {\it Markov chains and mixing times} (American Mathematical
Society, 2008).
\bibitem{Diaconis-96} P. Diaconis, The cutoff phenomenon in finite Markov chains, Proc. Natl. Acad. Sci. {\bf 93}, 1659 (1996).
\bibitem{Kastoryano-12} M. J. Kastoryano, D. Reeb, and M. M. Wolf, A cutoff phenomenon for quantum Markov chains, J. Phys. A: Math. Theor. {\bf 45}, 075307 (2012).
\bibitem{Kastoryano-13} M. J. Kastoryano and J. Eisert, Rapid mixing implies exponential decay of correlations, J. Math. Phys. {\bf 54}, 102201 (2013).
\bibitem{Vernier-20} E. Vernier, Mixing times and cutoffs in open quadratic fermionic systems, SciPost Phys. {\bf 9}, 049 (2020).

\bibitem{Lee-16} T. E. Lee, Anomalous Edge State in a Non-Hermitian Lattice, Phys. Rev. Lett. {\bf 116}, 133903 (2016).
\bibitem{Gong-18} Z. Gong, Y. Ashida, K. Kawabata, K. Takasan, S. Higashikawa, and M. Ueda, Topological Phases of Non-Hermitian Systems, Phys. Rev. X {\bf 8}, 031079 (2018).
\bibitem{Yao-18} S. Yao and Z. Wang, Edge States and Topological Invariants of Non-Hermitian Systems, Phys. Rev. Lett. {\bf 121}, 086803 (2018).
\bibitem{Kunst-18} F. K. Kunst, E. Edvardsson, J. C. Budich, and E. J. Bergholtz, Biorthogonal Bulk-Boundary Correspondence in Non-Hermitian Systems, Phys. Rev. Lett. {\bf 121}, 026808 (2018).
\bibitem{Thomale-18} C. H. Lee and R. Thomale, Anatomy of skin modes and topology in non-Hermitian systems, Phys. Rev. B {\bf 99}, 201103(R) (2019).
\bibitem{Song-19} F. Song, S. Yao, and Z. Wang, Non-Hermitian Skin Effect and Chiral Damping in Open Quantum Systems, Phys. Rev. Lett. {\bf 123}, 170401 (2019).
\bibitem{Borgnia-20} D. S. Borgnia, A. J. Kruchkov, and R. -J. Slager, Non-Hermitian Boundary Modes and Topology, Phys. Rev. Lett. {\bf 124}, 056802 (2020).
\bibitem{Okuma-20} N. Okuma, K. Kawabata, K. Shiozaki, and M. Sato, Topological Origin of Non-Hermitian Skin Effects, Phys. Rev. Lett. {\bf 124}, 086801 (2020).
\bibitem{Okugawa-20} R. Okugawa, R. Takahashi, and K. Yokomizo, Second-order topological non-Hermitian skin effects, Phys. Rev. B {\bf 102}, 241202(R) (2020).
\bibitem{Kawabata-20} K. Kawabata, M. Sato, and K. Shiozaki, Higher-order non-Hermitian skin effect, Phys. Rev. B {\bf 102}, 205118 (2020).

\bibitem{Daley-14} A. J. Daley, Quantum trajectories and open many-body quantum systems, Adv. Phys. {\bf 63}, 77 (2014).
\bibitem{Sieberer-16} L. M. Sieberer, M. Buchhold, and S. Diehl, Keldysh field theory for driven open quantum systems, Rep. Prog. Phys. {\bf 79}, 096001 (2016).

\bibitem{Lindblad-76} G. Lindblad, On the generators of quantum dynamical semigroups, Commun. Math. Phys. {\bf 48}, 119 (1976).
\bibitem{Gorini-76} V. Gorini, A. Kossakowski, and E. C. G. Sudarshan, Completely positive dynamical semigroups of $N$-level systems, J. Math. Phys. {\bf 17}, 821 (1976).

\bibitem{Eisler-11} V. Eisler, Crossover between ballistic and diffusive transport: the quantum exclusion process, J. Stat. Mech. (2011) P06007.
\bibitem{Temme-12} K. Temme, M. M. Wolf, and F. Verstraete, Stochastic exclusion processes versus coherent transport, New J. Phys. {\bf 14}, 075004 (2012).

\bibitem{Jaksch-03} D. Jaksch and P. Zoller, Creation of effective magnetic fields in optical lattices: the Hofstadter butterfly for cold neutral atoms, New J. Phys. {\bf 5}, 56 (2003).
\bibitem{Aidelsburger-13} M. Aidelsburger, M. Atala, M. Lohse, J. T. Barreiro, B. Paredes, and I. Bloch, Realization of the Hofstadter Hamiltonian with Ultracold Atoms in Optical Lattices, Phys. Rev. Lett. {\bf 111}, 185301 (2013).
\bibitem{Miyake-13} H. Miyake, G. A. Siviloglou, C. J. Kennedy, W. C. Burton, and W. Ketterle, Realizing the Harper Hamiltonian with Laser-Assisted Tunneling in Optical Lattices, Phys. Rev. Lett. {\bf 111}, 185302 (2013).
\bibitem{Dalibard-85} J. Dalibard and C. Cohen-Tannoudji, Atomic motion in laser light: connection between semiclassical and quantum descriptions, J. Phys. B {\bf 18}, 1661 (1985).
\bibitem{Pichler-10} H. Pichler, A. J. Daley, and P. Zoller, Nonequilibrium dynamics of bosonic atoms in optical lattices: Decoherence of many-body states due to spontaneous emission, Phys. Rev. A {\bf 82}, 063605 (2010).

\bibitem{Supplement} See Supplemental Material for details of the eigenvalue problem of the prototypical model, the relaxation dynamics of the off-diagonal elements of the density matrix, the physical implementation of the model, the effect of on-site dephasing on the Liouvillian skin effect, and the proof of the absence of the skin effect under the detailed balance condition.

\bibitem{Lieb-72} E. Lieb and D. Robinson, The finite group velocity of quantum spin systems, Commun. Math. Phys. {\bf 28}, 251 (1972).
\bibitem{Poulin-10} D. Poulin, Lieb-Robinson Bound and Locality for General Markovian Quantum Dynamics, Phys. Rev. Lett. {\bf 104}, 190401 (2010).

\bibitem{Hatano-96} N. Hatano and D. R. Nelson, Localization Transitions in Non-Hermitian Quantum Mechanics, Phys. Rev. Lett. {\bf 77}, 570 (1996).
\bibitem{Hatano-97} N. Hatano and D. R. Nelson, Vortex pinning and non-hermitian quantum mechanics, Phys. Rev. B {\bf 56}, 8651 (1997).
\bibitem{Hatano-98} N. Hatano and D. R. Nelson, Non-hermitian delocalization and eigenfunctions, Phys. Rev. B {\bf 58}, 8384 (1998).

\bibitem{Derrida-98} B. Derrida, An exactly soluble non-equilibrium system: The asymmetric simple exclusion process, Phys. Rep. {\bf 301}, 65 (1998).
\bibitem{Sandow-94} S. Sandow and G. Sch\"utz, On $U_q[SU(2)]$-Symmetric Driven Diffusion, Europhys. Lett. {\bf 26}, 7 (1994).

\bibitem{Chruscinski-10} D. Chru\'sci\'nski and A. Kossakowski, Non-Markovian Quantum Dynamics: Local versus Nonlocal, Phys. Rev. Lett. {\bf 104}, 070406 (2010).

\bibitem{Mori-20} T. Mori and T. Shirai, Resolving a Discrepancy between Liouvillian Gap and Relaxation Time in Boundary-Dissipated Quantum Many-Body Systems, Phys. Rev. Lett. {\bf 125}, 230604 (2020).

\bibitem{Chetrite-13} R. Chetrite and K. Mallick, Quantum fluctuation relations for the Lindblad master equation, J. Stat. Phys. {\bf 148}, 480 (2012).

\bibitem{Davies-79} E. B. Davies, {\it Quantum Theory of Open Systems} (IMA, London, 1979).

\bibitem{Spohn-78} H. Spohn and J. L. Lebowitz, Irreversible thermodynamics for quantum systems weakly coupled to thermal reservoirs, Adv. Chem. Phys. {\bf 38}, 109-142 (1978).

\bibitem{Kawabata-19} K. Kawabata, K. Shiozaki, M. Ueda, and M. Sato, Symmetry and Topology in Non-Hermitian Physics, Phys. Rev. X {\bf 9}, 041015 (2019).

\end{thebibliography}
\end{document}